\documentclass{article}
\usepackage{color}
\usepackage{authblk}
\usepackage{geometry}                
\usepackage{graphicx}
\usepackage{amssymb}
\usepackage{epstopdf}
\usepackage{soul}
\usepackage{amsmath,amsbsy}
\usepackage{cancel}
\usepackage[shortlabels]{enumitem}
\newcommand{\MM}[1]{\textcolor{black}{{#1}}}

\newcommand {\bs} {\mbox{\boldmath $s$}}
\newcommand {\bx} {\mbox{\boldmath $x$}}

\newcommand {\bp} {\bcancel{p}}
\newcommand {\bphi} {\bcancel{\phi}}

\newcommand {\bb} {\mbox{\boldmath $b$}}
\newcommand {\bc} {\mbox{\boldmath $c$}}

\begin{document}

\title{\textbf{Maximal Relevance and Optimal Learning Machines}}
\author[1]{O Duranthon}
\author[2]{M Marsili}
\author[3]{R Xie}
\affil[1]{{\small Department of Physics, \'Ecole Normale Sup\'erieure, 24 rue Lhomond, 75005 Paris, France}}
\affil[2]{{\small The Abdus Salam International Centre for Theoretical Physics, Strada Costiera 11, 34151 Trieste, Italy}}
\affil[3]{{\small Key Laboratory of Quark and Lepton Physics (MOE) and Institute of Particle Physics, Central China Normal University (CCNU), Wuhan, China}}
\date{}

\maketitle

\begin{abstract}%
We explore the hypothesis that learning machines extract representations of maximal relevance, where the relevance is defined as the entropy of the energy distribution of the internal representation. 
We show that the mutual information between the internal representation of a learning machine and the features that it extracts from the data is bounded from below by the relevance. This motivates our study of models with maximal relevance -- that we call Optimal Learning Machines  -- as candidates of maximally informative representations.
We analyse how the maximisation of the relevance is constrained both by the architecture of the model used and by the available data, in practical cases. We find that sub-extensive features that do not affect the thermodynamics of the model, may affect significantly learning performance, and that criticality enhances learning performance, but the existence of a critical point is not a necessary condition. On specific learning tasks, we find that {\em i)} the maximal values of the likelihood are achieved by models with maximal relevance, {\em ii)} internal representations approach the maximal relevance that can be achieved in a finite dataset and {\em iii)} learning is associated with a broadening of the spectrum of energy levels of the internal representation, in agreement with the maximum relevance hypothesis.
\end{abstract}

\section{Introduction}

Statistical mechanics models have been used in statistical learning since the pioneering works on associative memory \cite{Hopfield} and on Boltzmann Learning Machines \cite{BLM}. Many attempts to make sense of the spectacular performance of learning machines such as deep neural networks, have focused on understanding their statistical mechanics properties, as compared e.g. to spin glasses \cite{biroli}. 
Several attempts have been made to understand the typical properties of learning machines~\cite{tubiana2017emergence,decelle,zecchina,hennig}. 
These properties emerge after the interaction parameters have been learned from highly structured datasets \cite{mezard,goldt2019modelling,Gherardi2020}, and they depend on what measure of error or likelihood is used in supervised or unsupervised learning, respectively. 
This makes it hard to discuss the typical properties of learned models without reference to the particular dataset on which the model is trained, the objective function assumed, the algorithms or the architecture used. {This suggests that, in order to understand learning, we should focus on the ingredients which makes it possible~\cite{zdeborova2020understanding}.}


{This paper takes a different perspective, and it explores the claim of Refs.~\cite{SMJ,cjmrs} that learning maximises the {\em relevance} of the representation that the machine extracts from the data. Let us briefly recall the rationale behind this conclusion, which is articulated in Ref.~\cite{cjmrs}, to which we refer for more details.
In order to define the relevance, let us consider a generic \MM{unsupervised learning task that maps the data to the internal states of a learning machine, in order to capture the structure of statistical dependencies of the data}.
We assume that the data $\bx$ is generated from an unknown distribution $\bcancel{p}(\bx)$, that we shall call the {\em generative model}. 
A learning machine maps the data $\bx$ to the internal states $\bs$ of the hidden layer(s) of the learning machine. Training with data drawn from $\bp(\bx)$ induces a distribution $p(\bs)$ on the hidden states. 
We define the {\em energy} $E_{\bs}=-\log p(\bs)$ as the coding cost associated to a microscopic state $\bs$. The average energy
\begin{equation}
\label{ }
\langle E\rangle=-\sum_{\bs} p(\bs)\log p(\bs)\equiv H[\bs]
\end{equation}
coincides with the entropy of $\bs$, and it is a measure of the {\em resolution} of the representation. {Ref.~\cite{cjmrs} argues that the entropy of the energy }distribution
\begin{equation}
\label{relevance}
H[E]=-\sum_E p(E)\log p(E),\qquad p(E)=\sum_{\bs}p(\bs) \delta(E-E_{\bs})
\end{equation}
can be taken as a quantitative measure of the information that the model contains on the structure of the data, or of its generative model. In order to motivate this conclusion, we observe that $H[E]$ is typically small for Gibbs distributions, which encode states of maximal ignorance, because they have a very narrow distribution of energies. The internal representations $p(\bs)$ of a learning machine should be as far as possible from a maximum entropy model. Hence it should feature a broad spectrum of energies $E_{\bs}$, that corresponds to a large value of $H[E]$. In this representation, two states $\bs$ and $\bs'$ with very different energies $E_{\bs}$ and $E_{\bs'}$ necessarily code for significantly different patterns in the data, because they cannot occur typically, within the same maximum entropy model. 

Following Cubero {\em et al.}.~\cite{cjmrs}, we shall henceforth call $H[s]$ {\em resolution} and $H[E]$ {\em relevance}. 
The above discussion leads to the hypothesis that {\em machines that are trained on highly structured data should approach the limit where $H[E]$ is as large as possible}. In order to explore this claim, we define the class of Optimal Learning Machines (OLM) as the ideal limit of those statistical models that maximise the relevance at a given resolution $H[\bs]$
\begin{equation}
\label{maxrelgen}
\max_{\{E_{\bs}\}: H[s]=\bar E} H[E]
\end{equation}
over the distribution of energy levels $E_{\bs}$. The optimisation in Eq.~(\ref{maxrelgen}) is constrained, in practical cases, by the architecture of the machine. When these constrains are neglected, machines that satisfy Eq.~(\ref{maxrelgen}) have a broad distribution of energies with an exponential density of states~\cite{cjmrs}.
This, in turn, is equivalent to statistical criticality \cite{Mora,cjmrs}, which is a widely observed statistical regularity in efficient representations \cite{hennig,SMJ,MDL}. This corroborates the conjecture that the principle of maximal relevance applies to real learning machines. 

{The contributions of this paper is twofold: first we show that, in a generic learning task, the relevance lower bounds the mutual information between the internal representation and the features that the machine extracts from the data. This provides a rationale for why learning machines should approach the limit of maximal relevance, in order to extract the maximal amount of information from the data. 

Secondly, we investigate how the optimisation in Eq.~(\ref{maxrelgen}) is constrained both by the architecture of the machine used and by the amount of data on which it is trained. 
We start exploring the relevance of different architectures. Gaussian learning machines~\cite{karakida2016dynamical} offer an interesting counter-example, where the relevance $H[E]$ does not vary during training. This is consistent with the fact that the learned distribution $p(\bs)$ remains a Gaussian, irrespective of the generative model $\bp$. Hence, Gaussian learning machines do not learn anything on the generative model $\bp$, beyond its parameters. 

Taking $H[E]$ as a quantitative measure of learning performance, we study a class of toy learning machines with discrete energy levels, in order to characterise the architectural features associated with a high value of $H[E]$. Our results suggest that features that confer superior learning performance may be associated to sub-extensive features that are not accessible to standard statistical mechanics approaches. A superior learning performance is not necessarily related to the existence of a critical point separating two different phases, but when a critical point exists, learning performance improves when the model is tuned to its critical point.}

Finally, we present a series of numerical experiments, in order to understand how closely real learning machines approximate the ideal limit of OLM when the dataset is finite. Within the class of toy learning machines with discrete energy levels discussed earlier,we show that the models that have superior learning performance in two distinct (unsupervised learning) tasks are those with the highest values of $H[E]$. Then we move to Restricted Boltzmann Machines (RBM). There we argue that the number of samples on which the machine is trained limits the resolution of the energy scale. When the resolution and relevance are evaluated at the scale dictated by the available data, our numerical experiments show that RBM converge to states of maximal relevance, in agreement with~\cite{SMJ}. 
For small systems, it is possible to compute the full spectrum of energy levels and to follow its evolution. We show that during learning the distribution of energy levels indeed broadens. We summarise and comment our results in the concluding section.

\section{Relevance and hidden features}
\label{maxrel}


Let us assume that data points $\bx=(x_1\ldots, x_d)$ are generated as draws from an unknown distribution\footnote{
Here and in the rest of the paper, backslashed symbols (e.g. $\bp$) refer to unknown entities.} $\bp(\bx)$ that is characterised by a rich structure of dependence between the components $x_a$. 
For example, the MNIST dataset \cite{MNIST} of hand written digits has $d=28\times 28 = 784$ pixels, for each of which, the grey scale is codified in an integer $x_a$ in the range $[0,255]$. In spite of the fact that $\bx$ belongs to a very high-dimensional space, the MNIST dataset spans a manifold of only $d_{\rm int}\approx 13$ {\em intrinsic} dimensions~\cite{ansuini2019intrinsic}. This implies that $\bp(\bx)$ should be a model of strongly interacting variables, and that there should be a set of coordinates $\bphi(\bx)=(\bphi_1(\bx),\ldots,\bphi_{d_{\rm int}}(\bx))$ that describes the relevant variation in the dataset. We refer to $\bphi$ as the {\em hidden features}. \MM{As for $\bp$, the backslash indicates that $\bphi$ is an unknown theoretical construct that represents statistically significant patterns in the dataset}\footnote{In the case of the MNIST dataset, a draw from the generative model $\bp$ is a theoretical abstraction for the process of hand writing a digit by a human \MM{ and features are not necessarily related to ``meaningful'' patterns such as the values of digits. Deep learning machines also rely on features that look like noisy patterns, such as those used in adversarial examples. As shown by Ilyas {\em et al.}~\cite{ilyas2019adversarial}, 
machines that only use features which are reasonable for humans perform worse, in terms of generalisation, than those that use all features.} 
}. 

We focus on stochastic models, such as Restricted Boltzmann Machines (RBM) or Deep Belief Networks (DBN), with a finite number of discrete internal states. The goal of learning is to find a statistical model $p(\bs)$ over a discrete variable $\bs$, and a mapping $p(\bx|\bs)$ from $\bs$ to $\bx$, such that the generating distribution 
\begin{equation}
\label{pxdef}
p(\bx)=\sum_{\bs}p(\bx|\bs)p(\bs)
\end{equation}
is as close as possible to the empirical one (in unsupervised learning). The objective function employed may differ, yet in all cases, training aims at approximating the unknown generative process $p(\bx)\approx \bp(\bx)$ as closely as possible\footnote{In supervised learning the representation aims at reproducing a functional relation $\underline{x}_{\rm out}=f(\underline{x}_{\rm in})$ between two parts of the data $\bx=(\underline{x}_{\rm in},\underline{x}_{\rm out})$ as well as possible. This is similar to unsupervised learning with \hbox{$\bp(\bx)=\bp(\underline{x}_{\rm in})\delta({\underline{x}_{\rm out}-f(\underline{x}_{\rm in})})$}.}. Here we focus on the outcome of this process, abstracting from algorithmic details on how the representation is learned. We remark that, unlike $\bp$ which is a theoretical abstraction, $p(\bs)$ is a proper statistical model, as well as the induced distribution $p(\bx)$ in Eq.~(\ref{pxdef}). We define the energy of state $\bs$ as
\begin{equation}
\label{defE}
E_{\bs}=-\log p(\bs)
\end{equation}
which is the cost (in nats) for coding state $\bs$. 
\begin{figure}[ht]
\centering
\includegraphics[width=0.9\textwidth,angle=0]{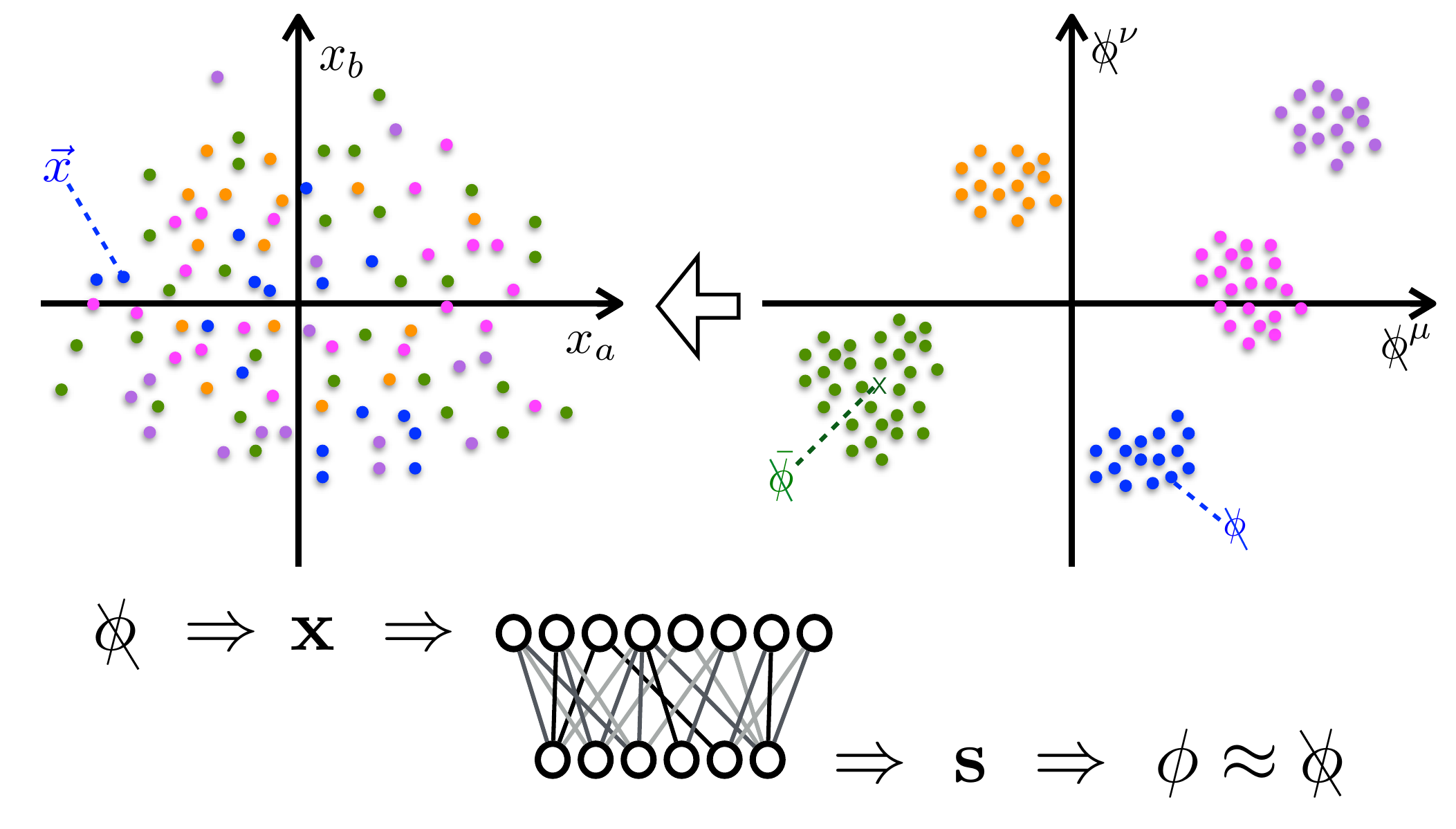}
\caption{\label{FigFeat} \MM{(Top) Sketch of the structure of statistical dependencies of the data in terms of hidden features~$\bphi$.} Points with similar features $\bphi\approx\bar{\bphi}$ are statistically identical. Statistically distinct points are distinguished by a significant variation of $\bphi$. \MM{(Bottom) Learning aims at representing this structure in the internal representation $\bs$ of a learning machine, in terms of extracted features~$\phi$ that approximately reproduce} the hidden features ($\phi\approx\bphi$).}
\end{figure}

If the learning machine captures correctly the structure of the data $\bx$ with which it has been trained, then it must {\em extract} features $\phi$ that approximate well the hidden features $\bphi$. Hence there must be a function $\phi(\bs)$ defined on the states of the hidden layers -- the {\em extracted} features -- such that its projection 
\begin{equation}
\label{phix}
\langle\phi|\bx\rangle\equiv\sum_{\bs} \phi(\bs)p(\bs|\bx)
\end{equation}
is close to the hidden features $\bphi(\bx)$. We call $\phi$ the {\em extracted} features, to contrast them with the hidden ones $\bphi$.

In an ideal situation, the distribution $p(\bs)$ should contain no further information than that contained in $\phi(\bs)$\MM{. In other words, $\phi$ should be the sufficient statistics for the distribution $p(\bs)$}. This means that the distribution of $\bs$ on all states with $\phi(\bs)=\bar\phi$ must be a maximum entropy distribution, i.e. $p(\bs|\bar\phi)={A(\bar\phi)}\delta_{\phi(\bs),\bar\phi}$, where $A(\bar\phi)$ is the inverse of the number of states $\bs$ with $\phi(\bs)=\bar\phi$. Put differently, states with the same features $\phi(\bs)=\phi(\bs')$ should be statistically indistinguishable, i.e. $p(\bs)=p(\bs')$ for all $\bs$ and $\bs'$ such that $\phi(\bs)=\phi(\bs')$. Therefore $E_{\bs}=E_{\bs'}$ for any two states $\bs$ and $\bs'$ such that $\phi(\bs)=\phi(\bs')$, which means that $E_{\bs}$ must be a function of $\phi(\bs)$. 
In other words, the statistical dependence of the variables can be represented as the Markov chain\footnote{The notation $z\to y\to w$ implies that, if $y$ is known, $w$ and $z$ are independent, i.e. that $I(z,w|y)=0$ \cite{CoverThomas}. This holds trivially true if $w$ is a function of $y$, because then $w$ is not a random variable, when $y$ is known.} 
\begin{equation}
\label{Markov}
\bs\to\phi\to E. 
\end{equation}
The data processing inequality \cite{CoverThomas} then implies that 
\begin{equation}
\label{dpi}
I(\bs,\phi)\ge I(\bs,E)=H[E],
\end{equation}
where $I(X,Y)$ is the mutual information between variables $X$ and $Y$ \cite{CoverThomas}, and the last equality comes from the fact that $I(\bs,E)=H[E]-H[E|\bs]$ and $H[E|\bs]=0$. Therefore, $H[E]$ provides a lower bound to the information $I(\bs,\phi)$ that the representation contains on the extracted features. Notice that $\phi$ is a function of $\bs$, and hence $I(\bs,\phi)=H[\phi]$. 


Going back to the hidden features $\bphi$, we observe that the training process can be represented by the Markov chain
\begin{equation}
\label{trainingMC}
\bphi\to\bx\to\bs\to\phi
\end{equation}
where $\bs$ is the representation that unfolds from training. \MM{If $\phi$ are sufficient statistics, which means that $\bx$ and $\bs$ are independent conditional on $\phi$, then the last step of the chain can be inverted, {\em i.e.} $\bphi\to\bx\to\phi\to\bs$. Then the}
data processing inequality implies that $I(\bs,\bphi)\le I(\bs,\phi)=H[\phi]$. Under the assumption that the variation $H[\phi]$ of the extracted features is entirely induced by the hidden features, it is reasonable to expect that this inequality is tight\footnote{If the last step of the Markov chain~(\ref{trainingMC}) can be reversed, then $I(\bs,\bphi)=I(\phi,\bphi)=H[\phi]-H[\phi|\bphi]=I(\bs,\phi)-H[\phi|\bphi]$. Therefore 
$I(\bs,\bphi)\approx I(\bs,\phi)$ is equivalent to the statement that the extracted features are approximately a function of the hidden features, i.e. $H[\phi|\bphi]\approx 0$.} 

The inequality (\ref{dpi}) is the main result of this Section. It connects the mutual information $I(\bs,\phi)$ of the extracted features with the relevance $H[E]$.
\MM{Apart from simple cases, such as the Gaussian learning machine discussed below, it is not clear how to identify the sufficient statistics of a distribution $p(\bs)$. So the} extracted feature $\phi$ are hardly accessible in practice, 
let alone the hidden ones $\bphi$. The distribution $p(\bs)$ instead can be computed for learning machines trained on structured data,  and hence $H[E]$ can be estimated, as we shall see in the following Sections.
The  inequality (\ref{dpi}) suggests that statistical models with a high value of $H[E]$ are natural candidates for good learning machines, even though there is no guarantee that the inequality (\ref{dpi}) is tight. 

The comparison between different learning machines on the basis of $H[E]$ makes sense only if the representations that they generate have the same resolution $H[\bs]$, i.e. the same coding cost.
For this reason, we shall consider the relevance $H[E]$ as a measure of the learning capacity, at a given level of resolution $H[\bs]$. An absolute maximum value for the maximum of $H[E]$ for a given $H[\bs]$ can be obtained by studying the problem
\begin{equation}
\label{maxHE}
\max_{\{W(E)\}:~H[\bs]} H[E]
\end{equation}
where $W(E)$ is the number of energy levels with $E_{\bs}=E$, which is constrained by the two conditions
\begin{eqnarray}
H[\bs] & = & -\sum_{\bs}p(\bs)\log p(\bs)=\sum_{E}W(E)e^{-E}E \\
N & = & \sum_{\bs} 1=\sum_{E}W(E)
\end{eqnarray}
where $N$ is the number of available states. This problem is studied in Appendix \ref{appILM}, where we show that the solution features and exponential density of states $W(E)=W_0e^{(1+\nu)E}$, where $\nu$ is a constant that depends on $H[\bs]$. 
This is an ideal limit, because in practical cases the maximisation is additionally constrained by the architecture of the learning machine and by the available data. 

Notice that $H[\bs|\phi]$ provides a measure of that part of $H[\bs]$ which is not informative on the features. By Eq. (\ref{dpi}), this is upper bounded by $H[\bs|E]$. Maximising $H[E]$ at fixed resolution $H[\bs]$ implies minimising $H[\bs|E]\ge H[\bs|\phi]$, i.e. squeezing noise out of the representation. 
An alternative definition of OLM, are machines that extract the most compressed representations from data, at a minimal information content $H[E]$ on the features
\begin{equation}
\label{ }
\min_{I(\bs,\phi)\ge H[E]}H[\bs].
\end{equation}

We remark that $H[E]$ is an unambiguous measure of information content when $E_{\bs}$ takes values in a discrete set. When $E_{\bs}$ can take any value on the real axis (as e.g. in RBMs), it is necessary to define the relevance with respect to a precision $\Delta$~\cite{CoverThomas}. In order to do this, we define the probability 
\begin{equation}
\label{pDelta}
p_\Delta(E)=\sum_{\bs: |E_{\bs}-E|\le\Delta/2} e^{-E_{\bs}}
\end{equation}
that a state $\bs$ has energy in the interval of width $\Delta$ around $E$. The relevance at precision $\Delta$ is defined as
\begin{equation}
\label{HEDelta}
H_\Delta[E]=-\sum_E p_\Delta(E)\log p_\Delta(E)
\end{equation}
where the sum on $E$ is restricted to integer multiples of $\Delta$. For a real random variable $E$ we expect that 
$H_\Delta[E]\simeq h[E]-\log\Delta$ for $\Delta\to 0$, where $h[E]$ is the differential entropy~\cite{CoverThomas}. 
We shall take $H_\Delta[E]$ as a measure of relevance, by making sure that $\Delta$ is chosen in the range where the relation $H_\Delta[E]\simeq h[E]-\log\Delta$ is satisfied.

\section{Exploring different architectures}

In this section, we explore classes of statistical models in order to understand how the relevance $H[E]$ depends on the architecture. Under the hypothesis that good learning machines are those with high values of $H[E]$, this sheds light on the architectural features that confer superior learning capacity to models. 

\subsection{The Gaussian learning machine and maximum entropy models}
\label{GaussianLM}

Consider a Gaussian learning machine with $m$ visible units $\bx=(x_1,\ldots, x_m)$ and $n$ hidden units $\bs=(s_1,\ldots,s_n)$ which are all continuous variables. The joint probability distribution is given by
\begin{equation}
\label{ }
p(\bx,\bs)=\frac{1}{Z(\bb,\hat w)}e^{-\frac 1 2 (\bx-\bb)^2 -  \frac 1 2 \bs^2 +\bx' \hat w \bs}
\end{equation}
where prime denote transpose, $\bb=(b_1,\ldots, b_m)$ is a vector of parameters, $\hat w$ is an $n\times m$ matrix of couplings, and $Z(\bb,\hat w)$ is a normalisation constant.
The couplings $\bb$ and $\hat w$ are adjusted so as to learn data generated from  $\bp(\bx)$. The marginal distribution of $\bx$ is given by a Gaussian
\begin{equation}
\label{data}
p(\bx)=\frac{(2\pi)^{n/2}}{Z(\bb,\hat w)}e^{-\frac 1 2 (\bx-\bb)^2 +\frac 1 2 \bx'\hat w\hat w '\bx}\,.
\end{equation}
{As shown in~\cite{karakida2016dynamical}, this machine learns the first two moments of the distribution of $\bx$. The dynamics of the parameters $\bb$ and $\hat w$ under different training algorithms has been discussed in Ref.~\cite{karakida2016dynamical} in detail.} In what follows, $\bb$ and $\hat w$ are the parameters of the trained machine, so they depend on the data. The distribution in the internal layer is\footnote{The largest singular value of $\hat w$ can be shown to be less than one, so that this is well defined.}
\begin{equation}
\label{ }
p(\bs)=\frac{(2\pi)^{m/2}}{Z(\bb,\hat w)}e^{-\frac 1 2 \bs' (1-\hat w'\hat w) \bs+\bb'\hat w\bs}
\end{equation}
so the energy is
\begin{eqnarray}
E(\bs) & = & \frac 1 2 \bs' (1-\hat w'\hat w) \bs-\bb'\hat w\bs -\frac{m}{2}\log (2\pi)+\log Z(\bb,\hat w) \\
 & = & E_0(\hat w)+\frac 1 2 (\bs-\bs_0)' (1-\hat w'\hat w) (\bs-\bs_0)
\end{eqnarray}
where 
\begin{equation}
\label{E0w}
E_0(\hat w)=\frac n 2 \log (2\pi) -\log\sqrt{{\rm det}(1-\hat w'\hat w)}
\end{equation}
is a constant (the ground state energy) and $\bs_0(\bb,\hat w)=(1-\hat w'\hat w)^{-1}\hat w\bb$ is a vector. 
Notice that the resolution is given by:
\begin{equation}
\label{Hsgauss}
H[\bs]=E_0(\hat w)+\frac n 2.
\end{equation}
The distribution of energies is
\begin{equation}
\label{ }
p(E)=\frac{1}{\Gamma(n/2)}\left(E-E_0\right)^{n/2-1}e^{-(E-E_0)}.
\end{equation}
Remarkably, $p(E)$ is independent of $\bb$ and it depends on $\hat w$, i.e. on the data, only through the constant $E_0$.
Hence the differential entropy \cite{CoverThomas} of this distribution 
\begin{equation}
\label{ }
h(E)=\frac n 2 +\log\Gamma\left(\frac n 2 \right)-\left(\frac n 2 -1\right)\log\psi \left(\frac n 2 \right)\simeq\frac 1 2 \log (4\pi e n)
+O(1/n)
\end{equation}
is independent of the parameters $\bb, \hat w$, i.e. it is independent of the data. The scaling $h(E)\simeq \frac 1 2 \log n$ for large $n$ is consistent with the fact that asymptotically the distribution of energies is a Gaussian with variance proportional to $n$. 


Let us now comment these results in terms of the discussion in Section \ref{maxrel}. In the Gaussian learning machine, the extracted features $\phi(\bs)=\{s_i,~i=1,\ldots,n,~s_i s_j, i\le j=1,\ldots,n\}$ coincide with the sufficient statistics. A Gaussian learning machine is a maximum entropy model of the form
\begin{equation}
\label{pscond}
p(\bs|\theta)=\frac{1}{Z(\theta)}e^{\theta \cdot\phi(\bs)},
\end{equation}
where the parameters $\theta=(\bb,\hat w)$ are adjusted so as to reproduce the means and covariances of the data. 
As a consequence, the generative model $p(\bx)$ is always Gaussian, irrespective of what $\bp$ is. A Gaussian learning machine does not learn anything on the {\em shape} of $\bp$, because it assumes its shape from the outset.
It only learns the parameters $\theta$. The amount of information learned on the features $I(\bs,\phi)$ equals $H[\bs]$, because $\bs$ is a part of $\phi$. 
Therefore, training increases the resolution $H[\bs]$ but it leaves the relevance $H[E]$ unchanged. 
The inequality (\ref{dpi}) $I(\bs,\phi)\ge H[E]$ remains true, but is not informative. 

An explicit manifestation of the poor performance of Gaussian learning machines, as compared e.g. to RBMs, in reproducing a dataset of handwritten digits, is given in Fig.~\ref{Fig:Sample}.
These findings suggest that the Gaussian multivariate model has a special place in statistical inference
 and the conclusions drawn from it should be taken with care. 

\begin{figure}[ht]
\centering
\includegraphics[width=0.95\textwidth,angle=0]{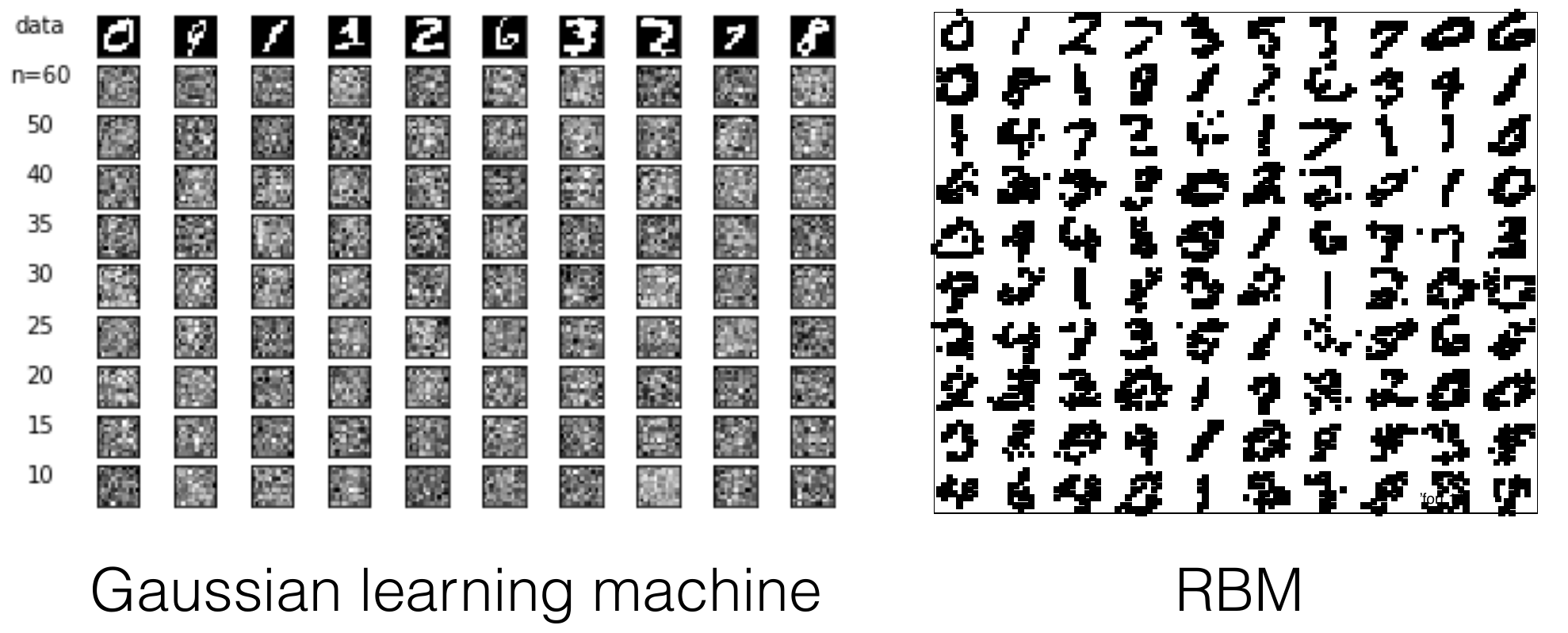}
\caption{\label{Fig:Sample} Comparison between sample digits of the (reduced) MNIST dataset (top row) and those generated by a Gaussian learning machine (left) or a RBM (right) with $n$ hidden nodes, for $n=10,\ldots,70$ (from bottom to top). Details on the dataset and on the RBM training experiments is provided in Section~\ref{sampleRBM}. }
\end{figure}
With respect to the picture in Fig.~\ref{FigFeat}, the Gaussian learning machine can only learn one point\footnote{Conversely, the expected value $\bar\phi$ over the whole dataset of $\phi(\bs)$ is sufficient to estimate the parameters of this model.} $\bar\phi$, and it cannot reproduce a wide variation of features, as in Fig.~\ref{FigFeat}.  

We expect the same to be generally true for exponential models of the type of Eq.~(\ref{pscond}). Indeed, for large $n$, the distribution of the energy $E_{\bs}$ is very narrow, because with very high probability
\[
E_{\bs}\approx \langle E \rangle_\theta=H[\bs].
\]
Schwab {\em et al.}~\cite{MSN} have shown that if the parameters $\theta$ themselves vary, as e.g. in
\begin{equation}
\label{pstheta}
p(\bs)=\int\!d\theta p(\bs|\theta)p(\theta).
\end{equation}
then the energy acquires a broad variation and it exhibits an exponential density of state\footnote{Ref.~\cite{cjmrs} shows that this result is a general consequence of the Asymptotic Equipartition Property in e.g. \cite{CoverThomas}).
If all typical states have the same probability $p(\bs|\theta)\sim e^{-nH[s]}$, then their number should be $W\sim e^{nH[s]}$.} $\log W(E)\propto E$. This is a defining feature of representations $p(\bs)$ that maximise the relevance. 
 In the present picture, the variation of the hyper-parameters $\theta$ is induced by the distribution $\bp(\bx)$ of the data $\bx$. In other words, the hidden features $\bphi$ play the role of the 
 ``hidden variables'' of Ref.~\cite{MSN}.
 
\subsection{Ising Learning Machines}

In this Section we shall consider a learning machines whose hidden variables $\bs$ are described by a fully connected model of Ising spins. These are maximum entropy models. So they are not particularly efficient as learning machines. Yet they provide a well defined playground for investigating the properties that make some architectures more efficient (i.e. with a larger $H[E]$) than others. This also allows us to contrast the properties of models with large $H[E]$ with those of well known models, such as the mean field Ising ferromagnet and the $J_{i,j}=\pm J$ spin glass. These insights make it possible to test whether models with the maximal $H[E]$ perform better than others in this class, on a specific application to real data. This will be the subject of Section \ref{secIsingtest}.

We consider learning machines with a distribution of internal states given by 
\begin{equation}
\label{model1}
p(\bs)=\frac{1}{Z}e^{-\mathcal{H}(\bs)},\qquad Z=\sum_s e^{-\mathcal{H}(\bs)},
\end{equation}
where $\bs=(s_1,\ldots,s_n),~s_i=\pm 1$ are $n$ spins. The Hamiltonian takes the form
\begin{equation}
\label{model}
\mathcal{H}(\bs)=-\sum_{i<j}J_{i,j}s_is_j,
\end{equation}
where the matrix $\hat J$ of couplings has elements $J_{i,j}\in\{\pm J\}$. 
The joint distribution  over the visible variables $\bx\in\{0,1\}^m$ and the internal state $\bs$ is given by 
$p(\bx,\bs)=p(\bx|\bs)p(\bs)$, where 
\begin{equation}
\label{pbxbsIsing}
p(\bx|\bs)=\frac{1}{Q_{\bs}}\exp \left[\sum_{j=1}^m  \left(b_j+\sum_{i=1}^ns_iw_{i,j}\right)x_j\right],
\qquad Q_{\bs}=\prod_{i=1}^m\left[
1+e^{b_j+\sum_i s_i w_{i,j} }\right].
\end{equation}
The parameters $\bb$, $\hat w$ and $J$ are learned from the data, e.g. by maximum likelihood maximisation. Here we concentrate on the distribution $p(\bs)$ of the internal variables $\bs$.

We define the energy as 
the coding cost $E_{\bs}=-\log p(\bs) =\mathcal{H}(\bs)+\log Z$, and address the optimisation problem
\begin{equation}
\label{Jstar}
\mathcal{J}={\rm arg}\max_{\hat J:~H[\bs]=\bar E}H[E], 
\end{equation}
where the maximum is taken over all matrices with elements $J_{i,j}=\pm J$, and on $J$. $\mathcal{J}$ is the subset of such matrices that achieve a maximal value of $H[E]$.

In order to compute $H[E]$ as a function of $\hat J$, 
we define the distribution of the energy as
\begin{equation}
\label{pE}
p(E)=\sum_{\bs} p(\bs)\delta(E-E_{\bs})=W(E)e^{-E}
\end{equation}
where $W(E)$ is the number of states $\bs$ with energy $E_{\bs}=E$. From this and Eq. (\ref{relevance}), we compute the value of $H[E]$. 



The rest of this section is devoted to describe the trade-off between $H[E]$ and $H[\bs]$ for models in the class of Eq. (\ref{model}). Since $H[\bs]\le n\log 2$ within this class, we shall discuss the rescaled resolution 
\begin{equation}
\label{ }
h_s=\frac{H[\bs]}{n\log 2}.
\end{equation}
Analogously, we observe that $(E_{\bs}-\log Z)/J$ takes values only on the integers $\left[-\frac{n(n-1)}{2},\frac{n(n-1)}{2}\right]$. This justifies the introduction of the rescaled relevance 
\begin{equation}
\label{ }
h_E=\frac{H[E]}{\log n}
\end{equation}
that takes values in the interval $[0,2)$. 
The sub-extensive nature of $H[E]$ suggests that properties that contribute to the relevance of a model may not be accessible to saddle point analysis of the partition function, that focuses only on the leading extensive terms. 
The behaviour of $H[E]$ on the number $n$ of spins depends on the fact that the model (\ref{model}) contains only one parameter $J$\footnote{For example, in a model with $\mathcal{H}(s)=-\sum_{i<j}J_{i,j}s_is_j-\sum_i h_i s_i$, $J_{i,j}=\pm J$ and $h_i=\pm h$, the energy takes values on a set of point $E=J n_J+h n_h$ with $n_J\in [-n(n-1)/2,n(n-1)/2]$ and $n_h\in [-n,n]$. So the number of possible values of $E$ grows at most as $n^3$, and we expect the $h_E\in [0,3)$. This suggests that $H[E]$ grows in general with $\log n$, with a coefficient that is bounded by a constant that grows linearly with the number of parameters.}. 

The information content of a model can be quantified in terms of the Stochastic Complexity in Minimum Description Length \cite{myung}. This in our case is given by 
\begin{equation}
\label{ }
{\rm SC}=\int\!\frac{dJ}{J}\sqrt{\mathbb{V}_J[E]},
\end{equation}
where $\mathbb{V}_J[E]$ is the variance of the distribution $p(E)$, 
%
which is closely related to the specific heat. 

A second measure of the information content of a model can be obtained by the area under the $h_E$ vs $h_s$ curve. This scores models according to their relevance across different levels of resolution. We shall call this measure Multi-Scale Relevance (MSR), following  \cite{CMR}, that introduced a similar measure on single samples.

Both $h_s$ and $h_E$ only depend on $\hat J$ through the degeneracy $W(E)$ of energy levels. $W(E)$ is invariant under the permutation of the spins. Hence if $\hat J\in\mathcal{J}$ then also the matrix with elements $J_{\pi(i),\pi(j)}$ is a solution, where $\pi(i)$ is a permutation of the indices $i=1,\ldots,n$. Also, matrices $\hat J$ and $\hat J'$ that are related by a gauge symmetry $J_{i,j}'=\tau_iJ_{i,j}\tau_j$, with $\tau_i=\pm 1$, have the same $W(E)$. Hence, if $\hat J\in\mathcal{J}$, then also its gauge transformed $\hat{J}'$ belongs to $\mathcal{J}$. We partially exploit these symmetries by fixing the gauge with the choice $J_{1,j}=1$ for all $j\neq 1$. Henceforth we shall focus on the reduced set $\mathcal{J}$ of matrices with this choice of the gauge. 

An interesting point that we shall discuss is how the set $\mathcal{J}$ of solutions $\hat J$ are organised, i.e. whether they correspond to isolated maxima or whether they form a connected set of nearby solutions in terms of local moves. 
We consider local moves that correspond to ``flipping'' one coupling $J_{\i,j}\to - J_{i,j}$, or that ``swap'' a pair of opposite couplings $J_{i,j}=-J_{k,l}$, changing their sign $(J_{i,j},J_{k,l})\to (-J_{i,j},-J_{k,l})$. Two models $\hat J$ and $\hat J'$ are connected if there is a sequence of such local moves that changes $\hat J$ into $\hat J'$. In this way, the set $\mathcal{J}$ can be divided into connected components. 
This analysis may shed light on the accessibility of solutions of Eq. (\ref{Jstar}) to local learning rules  during training, and it connects to the finding of wide flat minima~\cite{zecchina} in real learning machines\footnote{The latter refers to the fact that the landscape of the function that is optimised during training -- e.g. the error or (minus) the likelihood -- is characterised by wide flat minima in the space of parameters. If optimally trained machines correspond to models 
$\hat J$ of maximal relevance, then we expect that the set $\mathcal{J}$ has a large connected component.}. 

\subsubsection{Exact enumeration for small systems}
\label{sec:small}

Each model in Eq. (\ref{model}) is defined by a choice of the sign of the $n(n-1)/2$ couplings $J_{i,j}$ and by the strength $J$ of the couplings. 
There are $2^{n(n-1)/2}$ possible ways of choosing the signs. Yet, this number can be reduced by fixing the gauge. 
We were able to compute $h_E$ as a function of $h_s$ for all models up to $n=9$ spins. This allows us to find those models that achieve a maximal $h_E$ for a given resolution $h_s$.

\begin{figure}[ht]
\centering
\includegraphics[width=0.49\textwidth,angle=0]{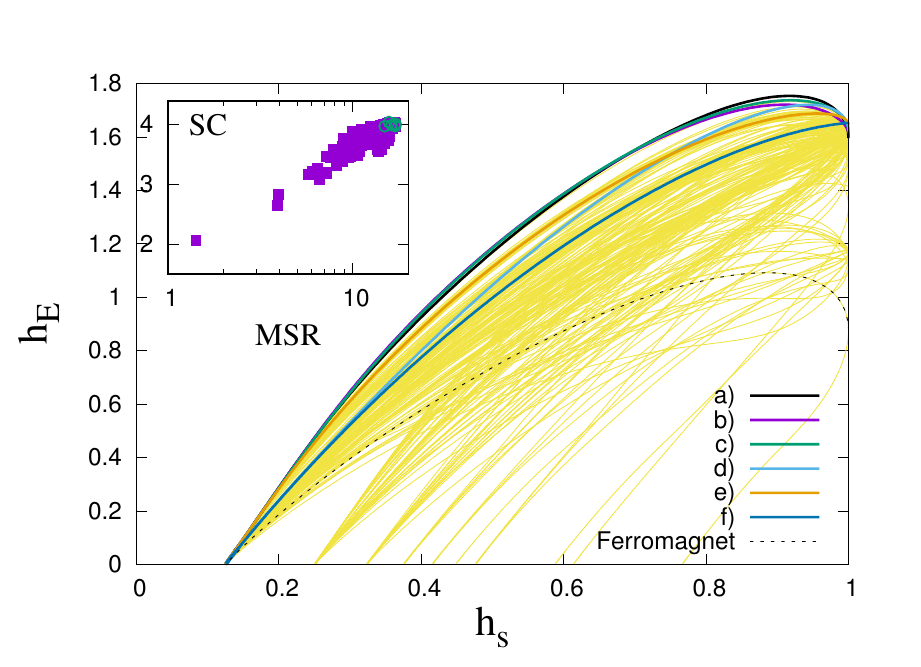}
\includegraphics[width=0.47\textwidth,angle=0]{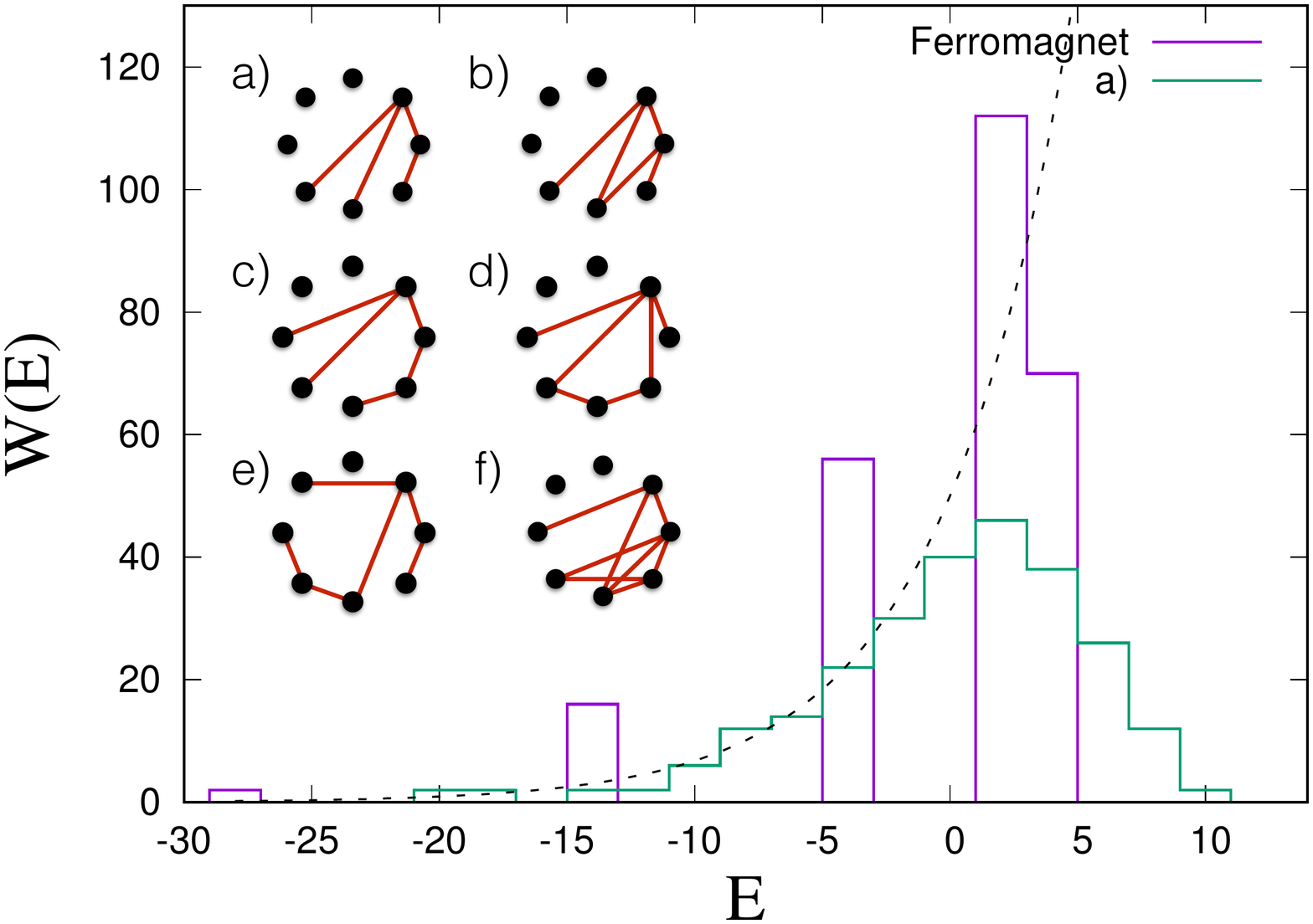}
\caption{\label{Fig1} (Left) $h_E$ vs $h_s$ for all models with $n=8$ spins. The curves highlighted correspond to the models shown on the right plot (only $J_{i,j}<0$ are shown as red links). The dotted line corresponds to the ferromagnet. The inset reports the relation between the stochastic complexity (SC) and the Multi-Scale Relevance (MSR), which is the area under the $h_E$ vs $h_s$ curve. The points corresponding to the models highlighted in the main plot are shown in green.
(Right) Degeneracy $W(E)$ of energy levels for the model {\em a)} (top left) and for the ferromagnet. The dotted line corresponds to an exponential $W(E)=W_0e^{\nu E}$ with $\nu=0.2$. }
\end{figure}

Representative results are shown in Fig. \ref{Fig1} for $n=8$. There are $M=219$ different curves in the left panel, each corresponds to a different degeneracy $W(E)$ of energy levels\footnote{For $n=3,4,5,6,7,8$ and $9$, we find $M=2,3,7,16,54,219$ and $1625$, respectively.}. 
Fig. \ref{Fig1}(left) shows, in particular, the six functions $h_E$ that achieve maximal relevance for some value of $h_s$. A representative choice of the corresponding matrix $\hat J\in\mathcal{J}$ is shown in the right part of Fig. \ref{Fig1}. We remark that there are many models  with the same $W(E)$, that cannot be reduced to one another by a permutation of the spins (for model {\em a} there are 325 different models). 
Fig. \ref{Fig1}(right) shows the degeneracy $W(E)$ for model {\em a)} (top left in Fig. \ref{Fig1} right). As a comparison, Fig. \ref{Fig1} also shows the curve $h_E(h_s)$ (left) and the degeneracy $W(E)$ for the ferromagnet ($J_{i,j}=J~\forall i,j$). 

The architectures that achieve maximal relevance, that we identify with OLM, have the following distinctive properties: {\em i)} the large majority of the couplings $J_{i,j}$ are positive, which means that OLM are close to ferromagnetic models. Indeed, the ground state of all OLM are the same as that of the ferromagnet. Yet, {\em ii)} the relevance for OLM achieves a much higher value than that of the ferromagnetic model (dotted line in Fig. \ref{Fig1} left). This is consistent with the fact that $W(E)$ for the ferromagnet is concentrated on few values of $E$, whereas for OLM it spreads over a larger number of values of $E$. For model {\em a)} $W(E)$ is consistent with an exponential behaviour, as predicted in \cite{cjmrs}. Finally {\em iii)} the architectures of OLM appear to be rather inhomogeneous, with negative $J_{i,j}$ impinging on a small subset of nodes. Such a low degree of symmetry, may be the origin of the fact that the number $|\mathcal{J}|$ of matrices $\hat J$ that share the same $W(E)$ of OLM is rather large (see Table \ref{tab2}). 


It is interesting to analyse the properties of the set $\mathcal{J}$ with respect to local dynamics, as discussed earlier. 
We find (see Table \ref{tab2} for $n=8$) that $\mathcal{J}$ is dominated by a single largest component, which is reminiscent of the wide flat minima discussed in \cite{zecchina}. 
As $h_s$ decreases, we see a tendency of the set to shrink and to fragment in different connected components. Yet, 
the union $\mathcal{J}_a\cup\mathcal{J}_b\cup\ldots\cup\mathcal{J}_f$ of all OLM $\in[a,b,\ldots,f]$, is composed of a single connected component.

Finally, the inset of Fig. \ref{Fig1} left) shows that models with a higher relevance, as measured by the MSR, typically have a larger Stochastic Complexity (SC).

\begin{table}
  \centering 
  \begin{tabular}{lrrr}
\hline
 OLM  & $|\mathcal{J}|$ & connected components & $h_s$ \\ \hline
 f  & 80640 & $80640\times 1$ & $[0.999,1]$ \\
 e & 40320 & $35280\times 1+5040\times 1$ & $[0.995,0.999]$  \\
 d & 10080 & $10080\times 1$ & $[0.984,0.995]$ \\
 a & 3360 & $2940\times 1+12\times 35$ & $[0.71,0.984]$  \\
 c & 10080 & $7560\times 1+1260\times 1+12\times 105$ & $[0.62,0.71]$ \\
 b & 3360 & $2100\times 1+420\times 2+4\times 105$ & $[0.22,0.62]$  \\
\hline
\end{tabular}
  \caption{Structure of the set of $\hat J$ corresponding to the OLM in Fig. \ref{Fig1} (right). The second column yields the size of the set $\mathcal{J}$ of matrices $\hat J$ that achieve maximal $h_E$ in a given interval ($4^{\rm th}$ column). The third column lists the connected components in $\mathcal{J}$ under rewiring of a single negative coupling. The format used is $\hbox{size}~\times~\hbox{multiplicity}~+~\ldots$. All models in the same class share the same (ferromagnetic) ground state.}\label{tab2}
\end{table}

\subsubsection{Some particular architectures for large $n$}
\label{largespinsect}

This section analyses specific architectures for which the curve $h_E(h_s)$ can be computed for moderately large values of $n$. The matrices $\hat J$ of the four models that we consider are schematically depicted in Fig.~\ref{FigFM} (left), together with the corresponding estimate of their capacity $h_E$. We discuss here only the main results and refer to the Appendix~\ref{AppSpin} for more details.

\paragraph{The mean field Ising ferromagnet and spin glasses}

The mean field Ising ferromagnet (MFIFM) corresponds to $J_{i,j}=J$ for all $i,j$ (model {\em a} in the left panel of Fig.~\ref{FigFM}). For large $n$, this is characterised by a high temperature disordered phase for $J<J_c/n$ and by a low temperature ordered phase for $J>J_c/n$ where a non-zero magnetisation spontaneously appears. The energy for the MFIFM can take at most $n+1$ different values, which gives an upper bound $h_E\le 1$. 
\begin{figure}[ht]
\centering
\includegraphics[width=0.45\textwidth,angle=0]{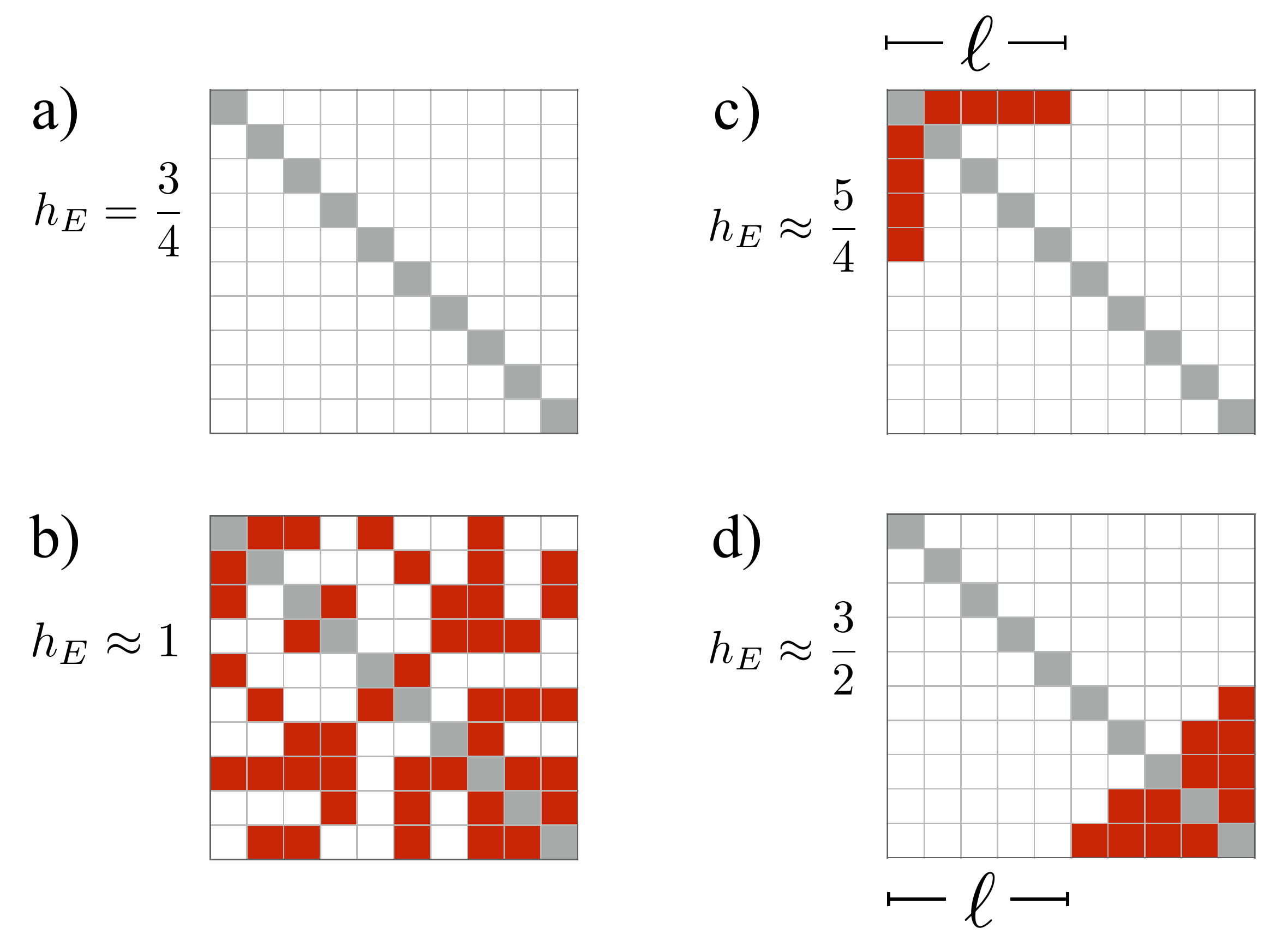}
\includegraphics[width=0.5\textwidth,angle=0]{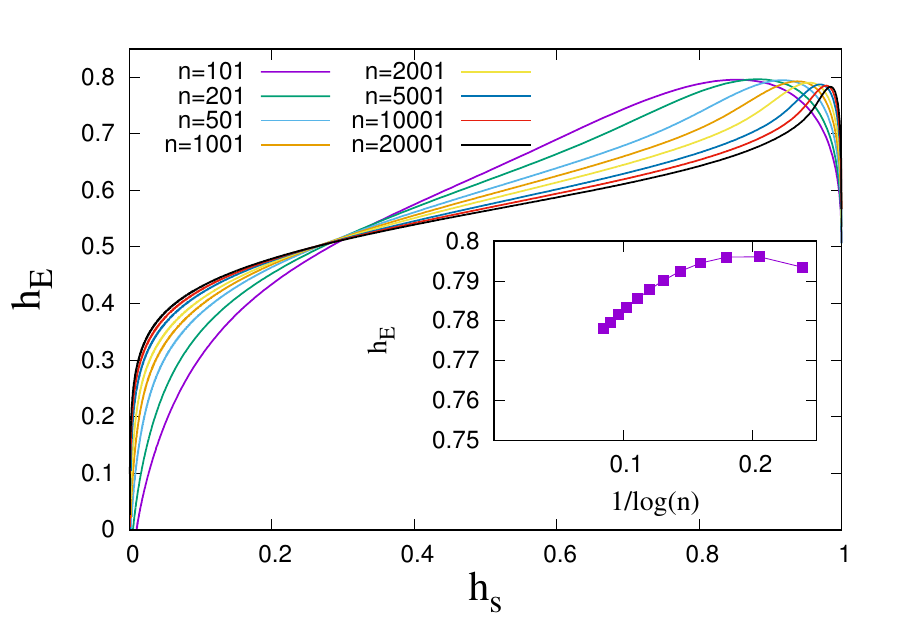}
\caption{\label{FigFM} (left) The coupling matrices $\hat J$ of the four models discussed in Section~\ref{largespinsect}: a) the MFIFM, b) the spin glass, c) the star model and d) the nested model. Red squares represent negative interactions. The maximal estimated value of $h_E$ is also reported for each model. (Right) $h_E$ vs $h_s$ for the MFIFM for different values of $n$. Inset: maximal value of $h_E$ as a function of $1/\log n$.}
\end{figure}

Fig. \ref{FigFM} (right) reports the curves $h_E(h_s)$ for several values of $n$. The curve exhibits a maximum in the neighbourhood of the critical point $J\approx J_c/n$. The decreasing part to the right of  the maximum corresponds to the high temperature phase, whereas the one to the left of the maximum to the low temperature phase. As $n$ increases, the maximum shifts to values of $h_s$ closer and closer to one. At the same time, the maximum gets sharper and sharper and its value slowly decreases towards a finite limit (see inset). 

The limiting form of the curve $h_E(h_s)$ can be computed in the limit $n\to\infty$. This shows that 
\begin{equation}
\label{ }
\lim_{n\to\infty} h_E=\left\{\begin{array}{cc}1/2 & J\neq J_c/n \\3/4 & J=J_c/n,\end{array}\right.
\end{equation}
which is consistent with the bound $h_E\le 1$. The analysis leading to this result (see Appendix \ref{AppSpin}) confirms that $h_E$ depends on sub-leading contributions that arise from the integration of fluctuations around the saddle point value.
At the critical point $J=J_c/n$, the resolution converges to $h_s\to 1$. Yet, depending on how the limit $h_s\to 1$ is taken, all values of $h_E\in[1/2,3/4]$ can be achieved. Indeed, for $J=0$ we have $h_s=1$ and $h_E=1/2$, exactly. Fig. \ref{FigFM} (right) is fully consistent with this result, although the convergence is very slow. 

When $J_{i,j}=\pm J$ are chosen at random with equal probability (model {\em b} in the left panel of Fig.~\ref{FigFM}), the model coincides with a spin glass. We know that the relevant scale for $J$ is $1/\sqrt{n}$. The ground state energy is extensive so $E_{\bs}$ spans a number of energy levels that is at most of order $n^{3/2}$. Therefore we expect that $h_E\le 3/2$ for a spin glass. In fact, for $J=c/\sqrt{n}$ with $c$ finite, the distribution of the energy extends over a range of order $\Delta E\propto \sqrt{n}$. Since the level spacing equals $2J\sim 1/\sqrt{n}$, this suggests that the number of energy levels on which $p(E)$ extends grows linearly with $n$. Hence we conjecture that $h_E\to 1$ as $n\to \infty$ for a model with randomly chosen $J_{i,j}$. This conclusion is corroborated by numerical results for systems of up to $n=28$ spins.

\paragraph{The star model}

Let us now consider a model that differs from a MFIFM by that fact that one spin has antiferromagnetic interactions with $\ell< n$ other spins (model {\em c} in the left panel of Fig.~\ref{FigFM}). 

\begin{figure}[ht]
\centering
\includegraphics[width=0.47\textwidth,angle=0]{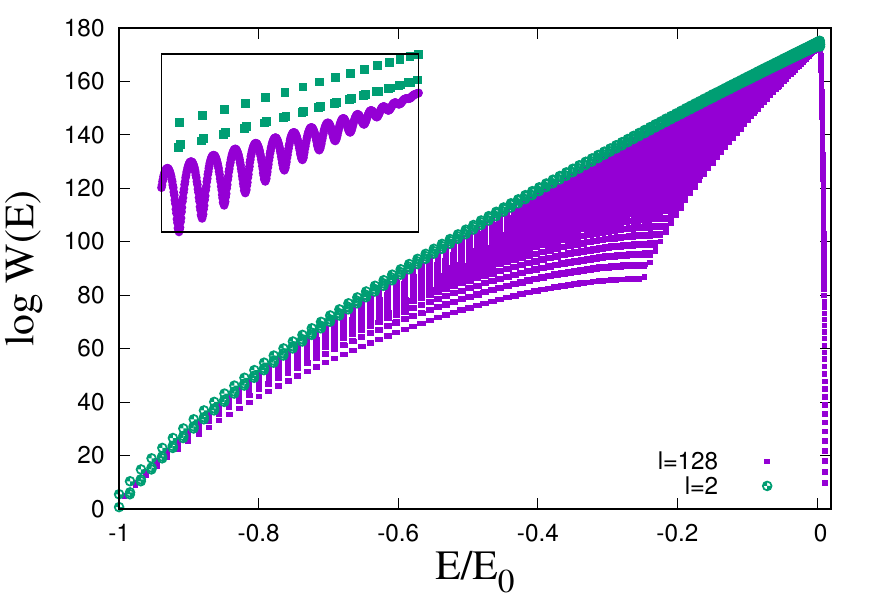}
\includegraphics[width=0.49\textwidth,angle=0]{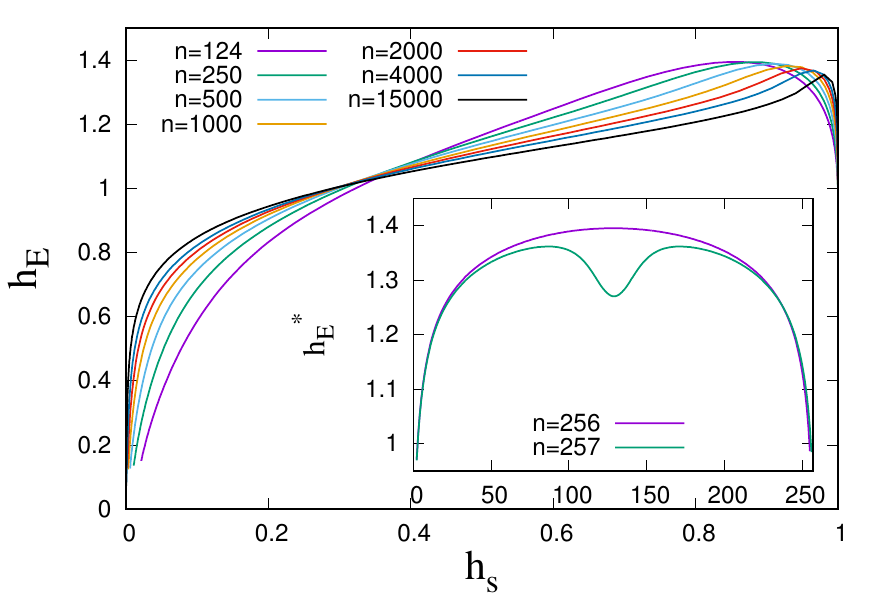}
\caption{\label{FigStar} (Left) Degeneracy of energy levels for $n=256$ and $\ell=2$ (green) and $128$ (violet). Energies are rescaled by $E_0=Jn(n-1)/2$. For $\ell=2$ there are $385$ distinct energy levels, whereas for $\ell =128$ there are $8511$. The inset magnifies the data in the range $E\in [-0.03E_0,0]$.
(Right) curves $h_E(h_s)$ for $\ell=n/2$ (full lines) and different values of $n$. Inset: maximal (over $h_s$) value of $h_E$ as a function of $\ell$, for $n=256$ and $257$.}
\end{figure}

The degeneracy $W(E)$ of energy levels is shown in Fig. \ref{FigStar}(left) for $n=256$, $\ell =2$ and $\ell=128$. As $\ell$ increases, the number of different values that $E$ takes also increases. For the MFIFM ($\ell=0$) $E$ can take only $n/2+1$ different values (for even $n$), whereas for $\ell=n/2$ we find that the number of possible values of $E$ grows almost as $n^2$. 
At the same time, $W(E)$ acquires a rapid variation as a function of $E$, so that for large $n$, $W(E)$ becomes a space filling curve (see Fig. \ref{FigStar} left and inset). Interestingly, the thermodynamics of the model is dominated by the convex envelope of $\log W(E)$, which is the same as that of the MFIFM. Therefore the star model and the MFIFM have indistinguishable thermodynamic properties. This is consistent with the fact that the number of $J_{i,j}=-1$ is a fraction of order $1/n$ of the total number of interactions and that they impinge on one out of the $n$ spins. It is easy to check that, for $\ell\le n/2$, the ground state of the star model is the ferromagnetic state $s_i=s_j$ for all $i\neq j$.

The maximal value of $h_E^*(\ell)=\max_{h_s}h_E$ for even values of $n$ is achieved at $\ell=n/2$ (see inset of Fig. \ref{FigStar} right), whereas for odd values of $n$, $h_E^*$ reaches a lower maximum at $\ell\approx n/3$. The symmetry for $\ell\to n-\ell$ of these curves is a consequence of the gauge transformation discussed above.
The curves $h_E(h_s)$ for $\ell=n/2$ and for different values of $n$ are shown in Fig. \ref{FigStar}(right). 
We find that the highest relevance $H[E]$ for even $n$ and $\ell=n/2$, grows faster than $\log n$. 
An extrapolation of the maximal value of $h_E$ is consistent with $h_E\to 5/4$ as $n\to\infty$. At fixed values of $h_s$ instead, we observe a slower growth of $H[E]$ with $\log n$, that is consistent with $h_E\to 1$ for large $n$. 
Note that the degeneracy of each energy level $E_m$ of the MFIFM is spread over $n$ other energy levels, $\sqrt{n}$ of which contribute to $H[E]$ (see inset of \ref{FigStar} left). This suggests a relation $h_E^{\rm Star}=h_E^{MFIFM}+1/2$, in agreement with numerical results.

Finally we remark that the set $\mathcal{J}$ of star models is connected in a single component, under a local dynamics updates of $\hat J$. 


\paragraph{The nested model}

In the next architecture that we consider, the couplings in the lower right corner of the matrix $\hat J$ are negative, and those on the diagonal or above are positive (model {\em d)} of Fig.~\ref{FigFM} left), i.e. $J_{i,j}={\rm sign} (n+\ell-i-j+1/2)$.
In these models, the set of negative $J_{i,j}$ is organised according to a nested network \cite{nested}.
Their properties are discussed in Appendix \ref{AppSpin} where we also derive a recursion relation for $W(E)$ that allows us to compute $h_E$ for moderately large values of~$n$. Fig. \ref{FigRec2} shows the results for different values of $n$ and for $\ell=1$ (left) and $n/2$ (right). 

\begin{figure}[ht]
\centering
\includegraphics[width=0.45\textwidth,angle=0]{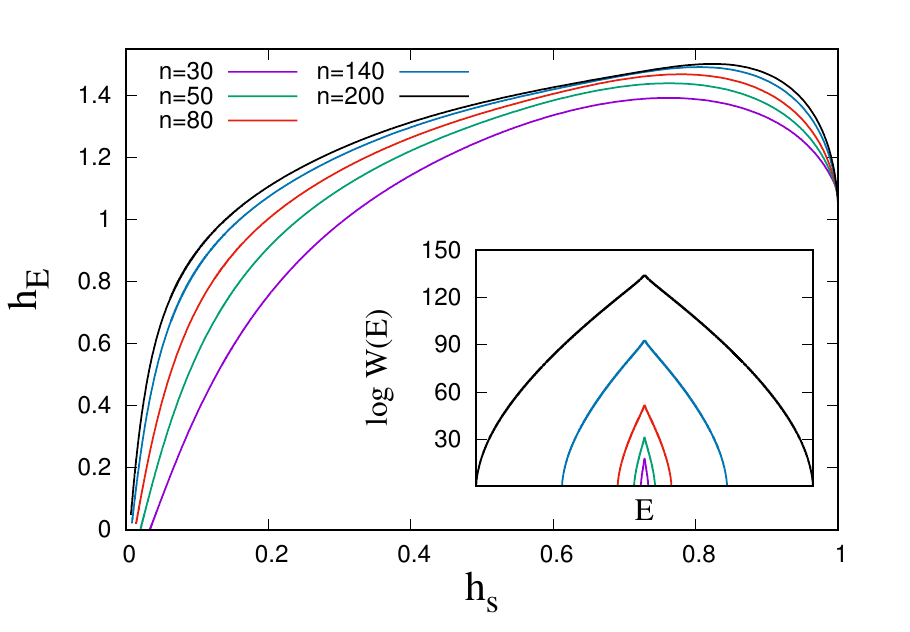}
\includegraphics[width=0.45\textwidth,angle=0]{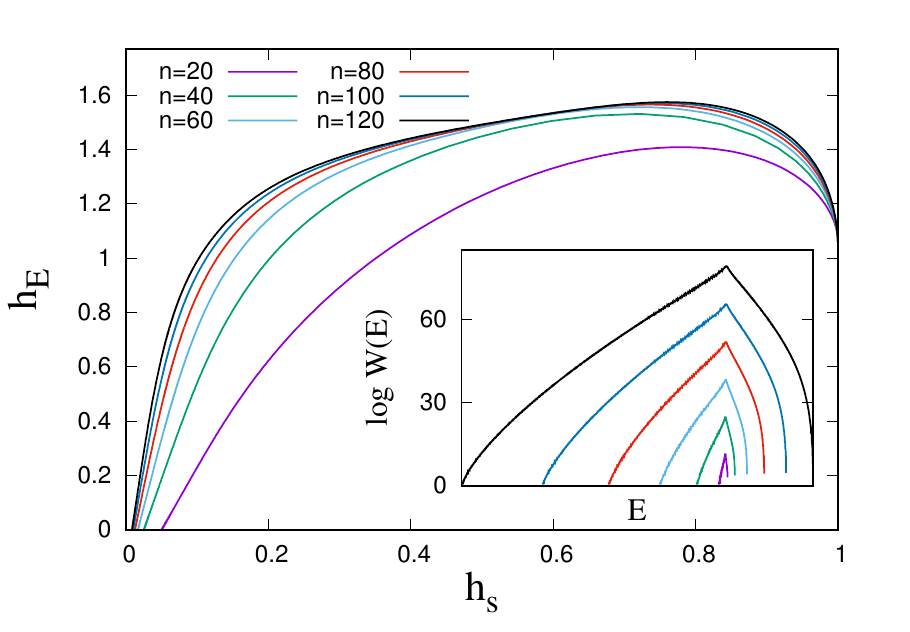}
\caption{\label{FigRec2} Relevance $h_E$ as a function of $h_s$ for the recursive model with $\ell=1$ (left) and $\ell=n/2$ (right),  for different values of $n$ (see legend). The inset shows the degeneracy $\log W(E)$ as a function of $E$ for the corresponding models.}
\end{figure}

Although we could not access values of $n$ as large as in the previous cases, Fig. \ref{FigRec2} shows that this model has qualitatively distinct features. First, we found that $\log W(E)$ approaches a continuous function of $E$ and it has support on a number of energy levels that increases as $n^2$. For $\ell=1$ we found that $W(E)=W(-E)$ is a symmetric function of $E$ (see inset of Fig. \ref{FigRec2} left). Second, in both cases, we found that $h_E$ increases monotonically with $n$ and it approaches a limiting value already for relatively small values of $n$. The maximal limiting values of $h_E$ is higher than that achieved with the other architectures ($h_E\approx 1.57$). We note that the number of anti-ferromagnetic interactions in these models is proportional to $n^2$, and hence the thermodynamic properties of nested models in the $n\to\infty$ limit differ from those of the MFIFM (even though the ground state is the same for $\ell\le n/2$).

There are $\frac{n!}{(n-\ell)!}$ possible models depending on how the set of $n-\ell$ fully ferromagnetic spins are chosen and on the ranking of the $\ell$ spins. This set of models forms a single connected component under local updates of the $J_{i,j}$. This is consistent with the conjecture that models with a high learning capacity are easily accessible by local updates of the couplings.

\section{Numerical experiments}

This section addresses the issue of whether the maximum relevance principle applies to learning experiments on real datasets. On the basis of this principle, we expect that {\em i)} models that achieve a higher value of $H[E]$ should describe better complex datasets, {\em ii)} within the precision allowed by a finite dataset, training converges to models of maximal relevance and {\em iii)} that learning induces broad distributions of energy levels. We address these three issues in turn.

\subsection{Complex data is best described  by models with maximal relevance}
\label{secIsingtest}

A prediction of the maximum relevance principle is that models that better fit a dataset are those with the highest relevance $H[E]$, provided the data contains a rich statistical structure. In order to test this prediction, we compared the ability of Ising models with different matrices $\hat J$ to represent a dataset, maximising the likelihood at different levels of resolution. 
 
Because of the exponential growth\footnote{To give an idea of the combinatorial complexity of the class of models, for $n=8$ there are $219$ different classes, each with many distinct models that cannot be reduced one to the other by a gauge transformation or by a permutation of the spins. For example, there are $325$ models with the same $H[E] - H[\bs]$ curve as model $a$ in Fig .\ref{Fig1}.} in the number of models with the number $n$ of spins, we limit our analysis to the case $n=5$. There are seven different classes of models (see Table \ref{Tabn5}). Each class contain models with the same $W(E)$ and hence with the same curve $h_E-h_s$. The class that achieves maximal value of $H[E]$ in a wide range of $H[\bs]$, that we shall call class A, contains two models\footnote{One with only one negative coupling and the other with three negative couplings incident on the same node, e.g. $J_{1,2}=J_{1,3}=J_{1,4}<0$.}. Class A also has the largest values of the SC and of the MSR. The other classes -- that we call B, C, D, E, FM  and AF -- contain one (E, AF and FM) or two models (B, C and D). Class FM contains the ferromagnetic model and AF the anti-ferromagnetic one.  

\begin{table}[htp]
\begin{center}\begin{tabular}{c|c|c|c} 
Model & MSR & SC & $\{(i,j):~J_{i,j}<0\}$\\ \hline
FM  & 4.176 &   2.768 & - \\
A  &  4.771 &     2.919 & $\{(1,2)\},~\{(1,2),(1,3),(1,4)\}$ \\
B  &  3.037 &     2.595 &    $\{(1,2),(1,3)\},~\{(1,2),(1,3),(2,3),(3,4)\}$ \\
C  & 1.510  &    2.154  & $\{(1,2),(1,3),(2,3)\},~\{(1,2),(1,3),(1,4),(2,3),(2,4)\}$ \\
D  & 3.656  &    2.861  & $\{(1,4),(2,3)\},~\{(1,2),(1,3),(1,4),(2,3))\}$ \\    
E  & 1.329  &    2.022  & $\{(1,2),(2,3),(3,4)\}$ \\
AF & 0.310 &   1.356   & all \\ \hline
\end{tabular} \caption{\label{Tabn5} The seven classes of models for $n=5$ spins. The last column lists the values $(i,j)$ for which $J_{i,j}<0$. Some of the classes (FM,AF and E) correspond to a single model, the others to two. The second and third column lists the values of the MSR and SC respectively.}
\end{center}
\label{defaulttable}
\end{table}


\begin{figure}[ht]
\centering
\includegraphics[width=0.8\textwidth,angle=0]{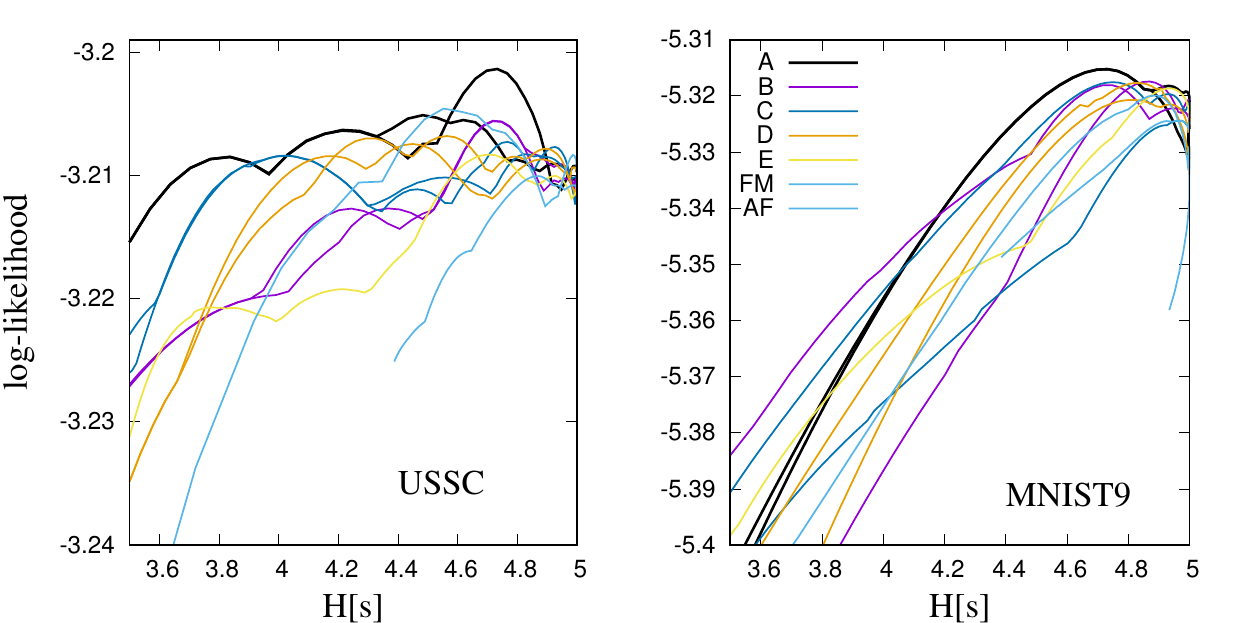}
\caption{\label{FigData} Log likelihood for the USSC (left) and MNIST9 data for models of $n=5$ spins, as a function of $H[\bs]$. 
Models are classified according to their $h_E-h_s$ curve (see Table \ref{Tabn5}).}
\end{figure}

We focus on two datasets of $m=9$ binary variables $\bx\in \{0,1\}^m$. One is that used in \cite{BialekUSSC} on the voting patterns in the US Supreme Court. This reports the votes (in favour or against) of the $m=9$ judges of the US Supreme Court on $N=895$ cases. The second is  obtained from the MNIST dataset~\cite{MNIST}, focusing on a patch of $3\times 3$ pixels of the $N=6\cdot 10^4$ hand written digits\footnote{Each MNIST data point contains $28\times 28$ pixels, with an integer value in the range $[0,255]$. We first transform the dataset into coarse grained pixels of $2\times 2$ original pixels, and binarise the result applying a threshold. Then we focus on the $3\times 3$ patch with pixels coordinates $(i,j)$ ranging from $5$ to $7$.}, as in \cite{hennig}. 
The joint distribution is built by taking $p(\bs)$ as in Eqs. (\ref{model1},\ref{model}) and $p(\bx|\bs)$ as in Eq.~(\ref{pbxbsIsing}). 
For every model and each value of $J$ we maximise the log-likelihood over the parameters $\bb$ and $\hat w$ by gradient ascent. Fig. \ref{FigData} shows that over a wide range of resolutions, the maximum of the likelihood is attained for models in class A. In particular, this interval contains the point where the log-likelihood achieves its maximum. 

\subsection{RBMs achieve the maximal relevance compatible with a dataset}
\label{sampleRBM}

In order to test whether learning machines converge to models of maximal relevance during training, we focus on the architecture of Restricted Boltzmann Machines (RBM). These have $n$ hidden nodes and $n_v$ visible nodes, where $n_v$ matches the dimensionality of the dataset. We study a reduced MNIST dataset with $n_v=144$ and $N=6\cdot 10^4$ data points. Details on RBMs and on the dataset are provided in Appendix~\ref{AppData}.

In general, the $2^n$ energy levels are learned from a sample of $N\ll 2^n$ data points. The finiteness of the data limits the precision with which the energy spectrum can be estimated. In order to take this into account, we observe that, with a change of variables $f=e^{-E}$, the differential entropies of $f$ and $E$ stand in the relation
\begin{equation}
\label{ }
h[E]=-\int_0^1\! df p(f)\log\left[p(f)\left|\frac{df}{dE}\right|\right]=h[f]+H[\bs]
\end{equation}
where $H[\bs]=\langle E\rangle$ is the average energy. The same relation holds for the variables at any precision $\Delta$, i.e. 
\begin{equation}
\label{HEsample}
H_{\Delta}[E]=H_{\Delta}[f]+H[\bs]
\end{equation}
The right hand side of Eq.(\ref{HEsample}) can be estimated in a finite sample $\hat \bs$, using the empirical distribution $\hat f_{\bs}=k_{\bs}/N$, where $k_{\bs}$ is the number of times that state $\bs$ occurs in the sample. 
Specifically, $H_{\Delta}[f]$ can be estimated using the empirical distribution of frequencies $P\{f_{\bs}=k/N\}=k m_k/N$, where $m_k$ is the number of states $\bs$ that occur $k$ times in the sample. Hence
\begin{equation}
\label{Hfsample}
H_{\Delta}[f]\simeq -\sum_{k}\frac{km_k}{N}\log \frac{km_k}{N}\equiv \hat H[k]\,.
\end{equation}
Likewise, the second term in Eq .(\ref{HEsample}) can be computed as 
\begin{equation}
\label{Hssample}
H[\bs] \simeq -\sum_{\bs}\frac{k_{\bs}}{N}\log\frac{k_{\bs}}{N} \equiv \hat H[\bs].
\end{equation}
The two quantities $\hat H[\bs]$ and $\hat H[k]$ have been introduced earlier as a measure of resolution and relevance, respectively, within a sample~\cite{MMR,HM,cjmrs}. 
Taken together, Eqs. (\ref{HEsample}, \ref{Hfsample}) and (\ref{Hssample}), imply
\begin{equation}
\label{HEsample1}
H_{\Delta}[E]\simeq\hat H[k]+\hat H[\bs]\,.
\end{equation}
We remark that the precision $\Delta$ is implicitly defined by the sample. The fact that both $\hat H[k]$ and $\hat H[\bs]$ are biased estimates of the entropy is immaterial for our purposes. We remind that our goal is that of establishing whether a learning machine approaches states of maximal relevance, for a given dataset. This can be done  comparing the values computed from a sample $\hat s$ drawn from $p(\bs)$, with the theoretical maximal value of $\hat H[k]+\hat H[\bs]$ that can be achieved in a sample of $N$ points. This has been computed in Ref,~\cite{HM}.

\begin{figure}[ht]
\centering
\includegraphics[width=0.95\textwidth,angle=0]{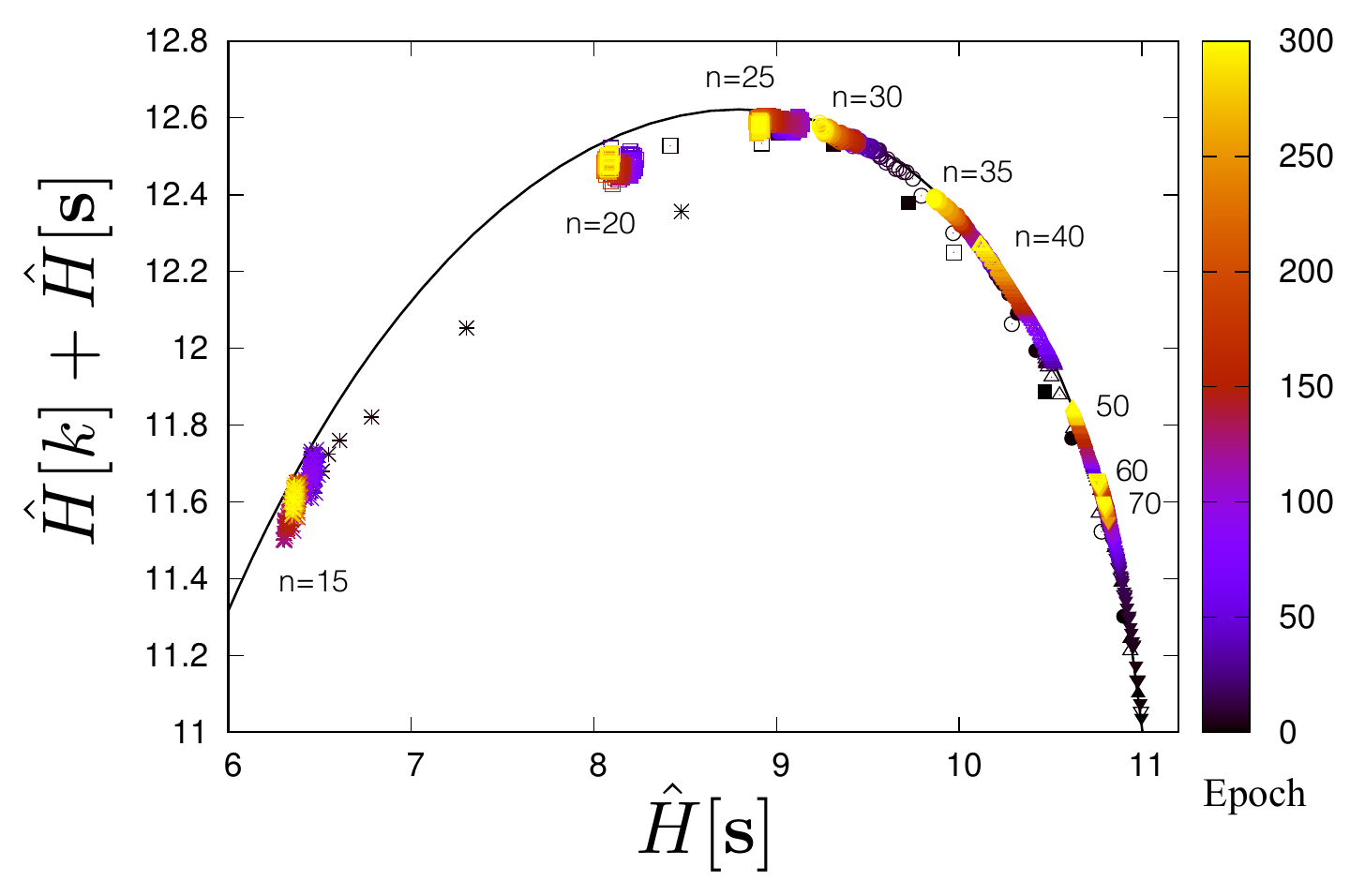}
\caption{\label{Fignha} Relevance $H_\Delta[E]\approx \hat H[k]+\hat H[\bs]$ as a function of $\hat H[\bs]\approx H[\bs]$, for different training sessions of RBMs' with $n=15,20,25,30,35,40,50,60$ and $70$ hidden nodes (from left to right). Each session was run using persistent contrastive divergence \cite{tieleman2008training} for $300$ epochs with mini batches of 10 samples. Sample digits drawn form $p(\bs)$ at the end of training for different $n$ are shown in Fig.~\ref{Fig:Sample}. }
\end{figure}

Fig. \ref{Fignha} reports the results of numerical experiments on the reduced MNIST dataset, where the number of hidden nodes of the RBM ranges from $n=15$ to $70$. Each training session lasts $300$ epochs with persistent contrastive divergence \cite{tieleman2008training}. 
Sample digits generated by the RBM are displayed in Fig.~\ref{Fig:Sample} (right). 
As a reference, the curve of maximal $\hat H[k]+\hat H[\bs]$~\cite{HM} is shown as a full black line. For each value of $n$, we observe trajectories that converge very fast to points that are close to the theoretical curve of maximal $\hat H[k]+\hat H[\bs ]$, as first observed in \cite{SMJ}. As learning proceeds, the internal state of the machine evolves towards more and more compressed representations (i.e. $\hat H[\bs]$ decreases).


\subsubsection{RBM develop broad distribution of energies}
\label{exactRBM}

Here we analyse the evolution of the energy distribution to test whether the distribution of energy broadens with learning, as predicted by relevance maximisation. The setting is the same of the previous section. We focus of RBM for small values of $n$, for which the distribution of the energy levels $E_{\bs}$, and hence  $H_\Delta[E]$, can be computed exactly during training. This allows us to monitor the evolution of the system as well as to appreciate the constraints imposed by the finiteness of the dataset and by the architecture. In order to do that, we take as a reference an ideal learning machine, that corresponds to the unconstrained maximisation of $H_\Delta [E]$, at a fixed $H[\bs]$. The way in which this is computed parallels Ref. \cite{cjmrs} and is detailed in Appendix \ref{appILM}. As discussed in \cite{cjmrs}, the distinguishing feature of an ideal learning machine is that the degeneracy $W(E)$ of energy levels follows an exponential behaviour $W(E)\simeq W_0 e^{\nu E}$. 

RBMs with $n\le 24$ hidden nodes have been trained with ($K=10$) contrastive divergence~\cite{hinton2012practical} on the data. We used mini-batches of $10^3$ samples for the results shown in Fig. \ref{FigRBMhE}. Although these are not ideal for learning\footnote{We confirmed that, as argued in \cite{hinton2012practical}, best performance is obtained with mini-batches of $\sim 10$ samples. Yet the trajectories obtained when the mini-batch size is small are very noisy.}, they allow us to monitor the dynamics of the RBM during learning, in the plane spanned by $H_\Delta[E]$ and $H[\bs]$.  
As a function of $\Delta$, the resolution is expected to behave as  $H_\Delta[E]=h[E]-\log\Delta$, where $h[E]$ is the differential entropy \cite{CoverThomas} of the probability density function of energy levels. Indeed we find that $H_\Delta[E]+\log\Delta$ is nearly constant in the range $0.1<\Delta<1$.

\begin{figure}[ht]
\centering
\includegraphics[width=0.95\textwidth,angle=0]{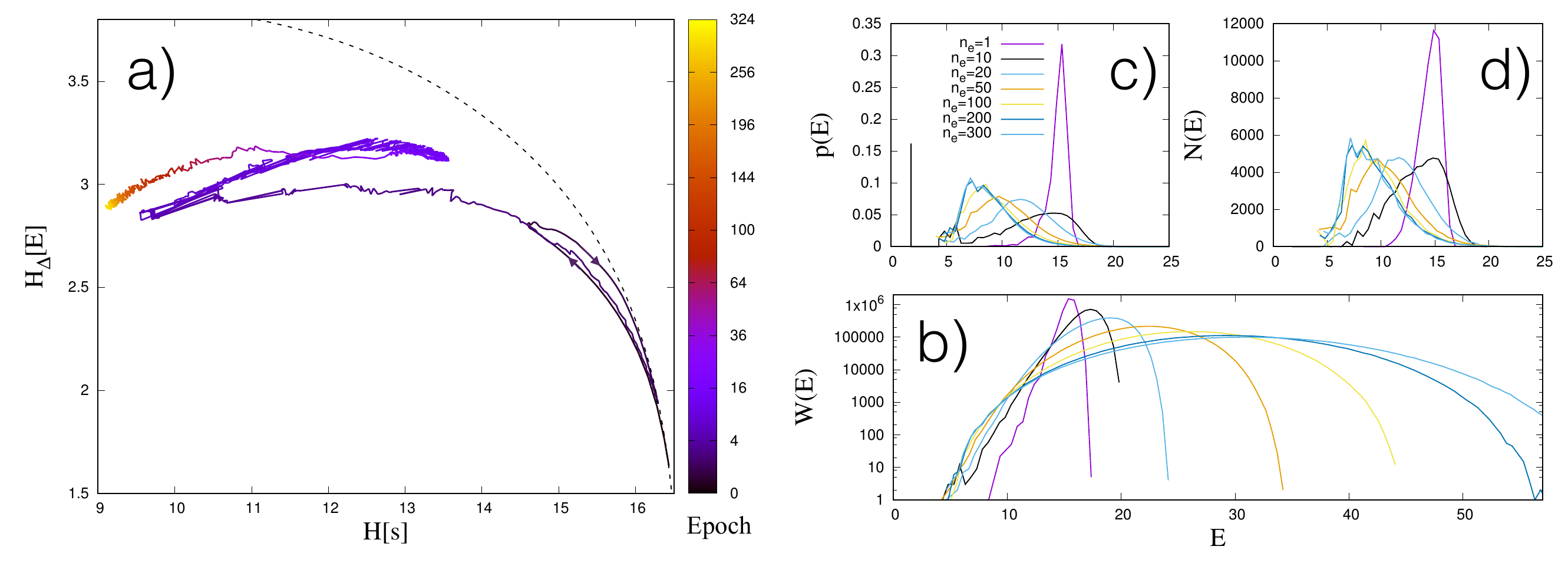}
\caption{\label{FigRBMhE} a) typical trajectory during learning in the plane $(H[\bs],H_\Delta[E])$ for a RBM with $n=24$ trained on the reduced MNIST data ($\Delta=0.5$). The ideal limit of maximal $H_\Delta[E]=h[E]-\log\Delta$ (see Appendix \ref{appILM}) is shown as a dashed line. b,c,d) Evolution of the distribution of energy levels for an 
RBM with $n=22$ hidden units, after $n_e=1,10,20,50,100,200$ and $300$ epochs. Panel b) shows the density $W(E)$ of energy levels, panel c) reports the probability density $p(E)=W(E)e^{-E}$. Panel d) shows the number $N(E)$ of data points that map to internal states of the machine with energy $E_{\bs}\in [E-\Delta/2,E+\Delta/2]$.}
\end{figure}

At different points in time, we compute the full spectrum of energy levels and compute $H_\Delta [E]$ and $H[\bs]$. Fig. \ref{FigRBMhE} (left) suggests that learning occurs in cycles, reminiscent of those discussed in Ref.~\cite{shamir2010learning}: A compression phase where $H[\bs]$ decreases, followed by an expansion phase where the system comes closer to the ideal limit of a machine with maximal $H_\Delta[E]$ (dashed line). 

Fig. \ref{FigRBMhE} (right) shows the evolution of the spectrum of energy levels. As the bottom panel (b) shows, the spectrum of energy levels expands more and more during training. At times ($n_e=1$, $n_e=10$) when the system is  closer to the ideal limit (dashed line in Fig. \ref{FigRBMhE} left), the degeneracy of energy levels follows approximately an exponential behaviour $W(E)\simeq W_0e^{\nu E}$, whereas it departs from it in the successive compression phases. In the late stages of the dynamics, for a given input $\bx$, the distribution $p(\bs|\bx)$ of internal states $\bs$ is sharply peaked. This makes it possible to map the dataset into a set of (clamped) internals states of the machine, as done in \cite{SMJ}, and to compute the corresponding distribution $N(E)$ of energies. As shown in the top right panel d) of Fig. \ref{FigRBMhE}, the evolution of this distribution during learning follows the one of $p(E)$ (panel c) and converges to a similar shape.

\section{Conclusion}

Maximal relevance has been proposed as a principle that characterises optimal learning machines \cite{cjmrs}. A distinctive feature of this approach is that it allows us to discuss the properties of learning machines without any reference to the data they have been trained with, as long as this contains a sufficiently rich structure. 

The first contribution of this paper is to provide a relation between the relevance and feature extraction in learning. We show that the relevance $H[E]$ provides a lower bound to the mutual information between the state of the machine and the features that a learning machine extracts (see Section \ref{maxrel}). This does not exclude the existence of efficient learning machines with low values of $H[E]$. Rather it ensures that learning machines with high values of $H[E]$ provide at least that much information on the features. 

The second contribution of this paper is to investigate architectures that achieve maximal relevance, and to confirm that, when the constraints imposed by the finiteness of the dataset and by the architecture are taken into account, the internal representations of well trained learning machines approach the limit of maximal relevance. 


We show that properties that confer a superior learning capacity, as measured by $H[E]$, may be  related to sub-extensive quantities in the statistical mechanics treatment. This suggests that the properties responsible for the spectacular performance of learning machines may not be accessible to approaches that focus only on leading thermodynamic terms. A proper understanding of these systems requires a detailed analysis of saddle point fluctuations. This result also tallies with the observation that complexity reveals itself in the sub-extensive contribution to the entropy, both in statistical models\footnote{Minimum Description Length, as well as Bayesian inference, identify the leading order term that should be used in model selection to penalise the likelihood of models for their complexity. In both cases this term is $\frac{d}{2}\log N$, where $d$ is the number of parameters of the model and $N$ is the number of data points. This term is sub-leading with respect to the log-likelihood, which is proportional to $N$.} \cite{schwarz1978,myung} and in time series\footnote{The complexity of a time series is defined in \cite{IlyaBialek} as the predictive information, which is the mutual information between the recent past, in a window of $T$ time points, and the future of the time series.} \cite{IlyaBialek}. 

For Ising systems, we find that local update rules can explore a large part of the set of OLM, suggesting that the relevance $H[E]$ is characterised by wide flat maxima in the space of the parameters. It is tempting to relate this property to the wide flat minima discussed by Baldassi {\em et al.}~\cite{zecchina}. 

Our results also contribute to the discussion on the criticality of learned models \cite{mastromatteo,hennig,statcrit}. We confirm that models have a superior learning performance when poised at the critical point. Yet, it is not strictly necessary for an OLM to be poised at a critical point marking the transition between two different phases. Indeed, we find recursive models with a learning performance superior to that of models that feature a phase transition, in a wide range of the resolution scales. 

We hope that these results will not only contribute to our understanding of learning machines, but that they may also pave the way to applications that may improve further the performance of learning machines. 

\section*{Acknowledgments}
We benefitted from discussions with Jean Barbier, Riccardo Zecchina, Federico Ricci-Tersenghi and Yasser Roudi. We are grateful to the authors of Re. \cite{BialekUSSC} for sharing their data with us. We thank an anonymous referee of MSML2020 for pointing out the example in Section \ref{GaussianLM}.

\appendix

\section{The ideal limit of an Optimal Learning Machine}
\label{appILM}

For a  learning machine with many hidden units ($n\gg 1$), we can approximate the distribution of energies with a continuous probability density $p(E)$, that we assumed to be defined in an interval $E\in [0,\bar E]$. 
For $\Delta$ small enough, the behaviour of $H_\Delta[E]$ with $\Delta$ is given by 
\begin{equation}
\label{ }
H_\Delta [E]\simeq h[E]-\log\Delta
\end{equation}
where $h[E]$ is the differential entropy~\cite{CoverThomas}
\begin{equation}
\label{ }
h[E]=-\int_0^{\bar E}\! dE p(E)\log p(E).
\end{equation}
For a fixed $\Delta$, the ideal limit of a learning machine with maximal $H_\Delta[E]$ is then obtained by finding the $p(E)$ that makes $h[E]$ maximal.

Let $W(E)$ be the density of energy levels, i.e. $W(E)dE$ is the number of states $s$ with $E_s\in [E,E+dE)$. The probability density function $p(E)$ is then given by $p(E)=W(E)e^{-E}$. When no other constraint is imposed, Optimal Learning Machines are defined by the solution of the optimisation problem
\begin{eqnarray}
W^*(E) & = & {\rm arg}\max_{W(E)} h[E] \\
 & ~ & \hbox{s.t.}~~~H[\bs]=\langle E\rangle \label{constE}\\
 & ~ & \qquad \int_0^{\bar E} W(E)dE=N\label{constN}\\ 
 & ~ & \qquad \int_0^{\bar E} W(E)e^{-E}dE=1,\label{const1}
\end{eqnarray}
where
\begin{equation}
\label{ }
h[E]=-\int_0^{\bar E}\! dEp(E)\log p(E)=\langle E\rangle-\langle\log W(E)\rangle\,,
\end{equation}
and $N$ is the total number of states $\bs$.
The solution of this problem is found introducing Lagrange multipliers to enforce the constraints (\ref{constE},\ref{const1}), and fixing $\bar E$ to satisfy Eq.  (\ref{constN}). The solution reads
\begin{eqnarray}
W(E) & = & \frac{\nu}{e^{\nu\bar E}-1}e^{(1+\nu)\bar E}, \\
H[\bs] & = & \langle E\rangle = \frac{\bar E}{1-e^{-\nu\bar E}}-\frac{1}{\nu},\label{eqE}\\
h[E] & = & \log\frac{e^{\nu \bar E}-1}{\nu}-\frac{\nu\bar E}{1-e^{-\nu\bar E}}+1,\\
N & = & \frac{\nu}{1+\nu}\frac{e^{(1+\nu)\bar E}-1}{e^{\nu\bar E}-1} \label{eqN},
\end{eqnarray}
where Eq. (\ref{eqN}) needs to be solved to obtain $\bar E$ as a function of $N$ and 
Eq. (\ref{eqE}) yields $\nu$ as a function of $H[\bs]$. The limit $\nu\to 0$ is revealing to understand the scaling of $H[\bs]$ and $h[E]$ on $N$. We find
\begin{eqnarray}
\bar E & = & \log (1+N\bar E)\simeq \log N+\log\log N+O\left(\frac{\log\log N}{\log N}\right)\qquad (\nu=1)\\
H[\bs] & = & \frac 1 2 \bar E\simeq \frac 1 2 \log N\\
h[E] & = &  \log \bar E\simeq \log \log N.\label{maxhE}
\end{eqnarray}
At this point, the curve $h[E]$ vs $H[\bs]$ is close to its maximum. So Eq. (\ref{maxhE}) provides an upper bound on the possible value of $h[E]$. In the case of a Boltzmann machine with $n$ hidden binary units, there are $N=2^n$ hidden states. Hence $H[\bs]$ grows linearly with $n$ and $H[E]\propto \log n$ grows with the logarithm of the number of hidden units $n$. This is the same scaling that we have found for discrete energy levels, with $\Delta=J$.


\section{Spin models}
\label{AppSpin}

\paragraph{The MFIFM model}

The energy for the MFIFM ranges over an interval $[-Jn(n-1)/2,Jn(n-1)/2]$ of order $n^2$, but energies $E_{\bs}$ can take at most $n+1$ different values 
\begin{equation}
\label{EmMFIFM}
E_m=-\frac{J}{2}(m^2-n),\qquad m=\sum_{i=1}^ns_i=-n,-n+2,\ldots,n-2,n.
\end{equation}
This implies that $h_E\le \log(n+1)/\log n$. Note also that there is a single matrix $\hat J$ which corresponds to the degeneracy of energy levels $W(E)$ of the MFIFM (apart from gauge transformations)..

The distribution of energy levels is given by $p(E)=W(E)e^{-E}/Z$, where $Z=\sum_E W(E)e^{-E}$. Hence the relevance is given by
\begin{equation}
\label{eqHEMFIFM}
H[E]= \log Z -\langle\log W(E)\rangle+\langle E\rangle.
\end{equation}
The degeneracy of energy level $E_m$ in Eq. (\ref{EmMFIFM}) is 
\[
W(E_m)=2{n\choose \frac{n+m}{2}}\simeq
\sqrt{\frac{2}{\pi n(1-\mu^2)}}
e^{n S(\mu)},\qquad \mu=\frac{m}{n}\in [-1,1]
\]
where $S(\mu)=-\frac{1+\mu}{2}\log\frac{1+\mu}{2}-\frac{1-\mu}{2}\log\frac{1-\mu}{2}$. In the regime where $h_s\in [0,1]$ is finite, $J$ is of order $1/n$ and, for $n\to\infty$, the partition function is dominated by the saddle point 
$\mu^*={\rm arg}\min_\mu f(\mu)$ with $f(\mu)=Jn\mu^2/2-S(\mu)$. 
For $Jn\neq 1$, the first term in Eq. (\ref{eqHEMFIFM}) can be computed with integration over the gaussian fluctuations 
\[
Z\simeq \sqrt{\frac{2n}{\pi}} \int_{-1}^1 \! \frac{d\mu}{\sqrt{1-\mu^2}} e^{-nf(\mu)}\simeq \sqrt{\frac{2}{\pi (1-{\mu^*}^2) f"(\mu^*)}}e^{-nf(\mu^*)}
\]
Here, the $\sqrt{n}$ factor is canceled because $f(\mu)-f(\mu^*)\sim (\mu-\mu^*)^2$, and the change of variables $z=\sqrt{n}(\mu-\mu^*)$ generates a $1/\sqrt{n}$ term.

Hence $\log Z\simeq -nf(\mu^*)+{\rm const}$. The extensive term in $\log Z$ is canceled by an analogous term in
\[
\langle\log W(E)\rangle-\langle E\rangle\simeq n f(\mu^*)+\log\sqrt{2\pi n(1-\mu^2)}
\]
so, to leading order, $H[E]\simeq \frac{1}{2}\log n+{\rm const}$, which implies $h_E\to 1/2$ as $n\to\infty$.

For $Jn=1$, instead, $f(\mu)-f(\mu^*)\simeq a(\mu-\mu^*)^4+\ldots$. This implies that in the calculation of $Z$ we need a change of variables $z=n^{1/4}(\mu-\mu^*)$ that yields $\log Z\simeq \frac{1}{4}\log n+{\rm const}$. This additional term is responsible for the asymptotic behaviour $h_E\to 3/4$ for $Jn=1$.

\paragraph{The spin glass model}

We computed numerically the value of $\max_J h_E$ for $100$ realisations of models with random $J_{i,j}=\pm 1$ (with $P\{J_{i,j}=+1\}=1/2$), for systems of up to $n=28$ spins. A linear fit of $H[E]$ versus $\log n$ yields $h_E\simeq 1.02 \pm 0.01$. 

\paragraph{The star model}

In the start model we take $J_{i,j}=-1$ if $i=1$ and $j\le \ell+1$, and $J_{i,j}=+1$ otherwise. We divide the $n-1$ group of spins, except spin $s_1$, into two groups: the $\ell$ ones with anti-ferromagnetic interaction with $s_1$ and the remaining $n-\ell-1$. If $q$ and $k$ are the number of positive spins in the first and second group, respectively, the energy is given by
\begin{equation}
\label{ }
\epsilon(k,q,s_1) = J\left[s_1(n-1-2\ell-2k+2q)-\frac{1}{2}\left(2k+2q-n+1\right)^2+\frac{n-1}{2}\right].
\end{equation}
The degeneracy of energy levels is given by
\begin{equation}
W(E) = \sum_{k=1}^{n-1-\ell}\sum_{q=1}^\ell\sum_{s_1=\pm 1}{\ell\choose q}{n-1-\ell\choose k}\delta_{E,\epsilon(k,q,s_1)}.
\end{equation}
Notice that for $\ell=0$ one recovers the MFIFM. A gauge transformation $J_{i,j}\to \tau_i J_{i,j}\tau_j$, with $\tau_1=-1$ and $\tau_i=+1$ for $i>1$, maps a model with  $\ell>n/2$ into a model with $\ell'=n-\ell\le n/2$. So it is sufficient to study the model for $\ell\le n/2$.

Each models $\mathcal{J}$ with the star structure can be ``reached'' from any other model, under a dynamics where a single $J_{i,j}\mapsto -J_{i,j}$ changes sign. If we consider a ``rewiring'' dynamics where the signs of a pair $(J_{i,j},J_{i,k})=(-1,+1)$ of links are swapped $(J_{i,j},J_{i,k})\mapsto (-J_{i,j},-J_{i,k})$, each of the $\ell$ negative links can be rewired to any of the $n-\ell-1$ spins with all ferromagnetic interactions. This dynamics, spans the whole subset of models in $\mathcal{J}$ with the same hub node. These subsets are related by a permutation of the nodes.

\paragraph{The nested model}

The triangle model has 
\[
J_{i,j}=\left\{\begin{array}{cc} -1 & \hbox{for}~ i>\ell,~j> n+\ell-i\\
+1 & \hbox{otherwise}\end{array}\right.
\]
where $\ell=1,\ldots,n-1$. For $\ell=1$ this corresponds to a matrix $\hat J$ where all the elements below the diagonal $i+j=n+1$ are negative, as shown in Fig. \ref{FigRec}(left). 
In these models, the set of negative $J_{i,j}$ is organised according to a nested network \cite{nested}.

\begin{figure}[ht]
\centering
\includegraphics[width=0.3\textwidth,angle=0]{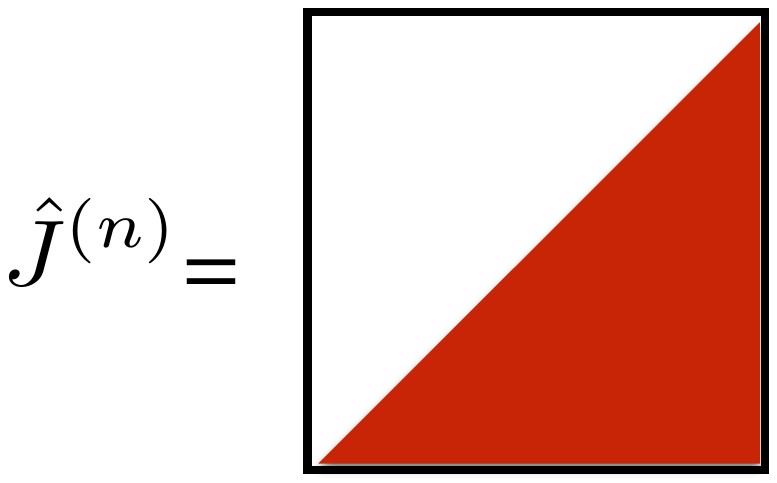}
\includegraphics[width=0.5\textwidth,angle=0]{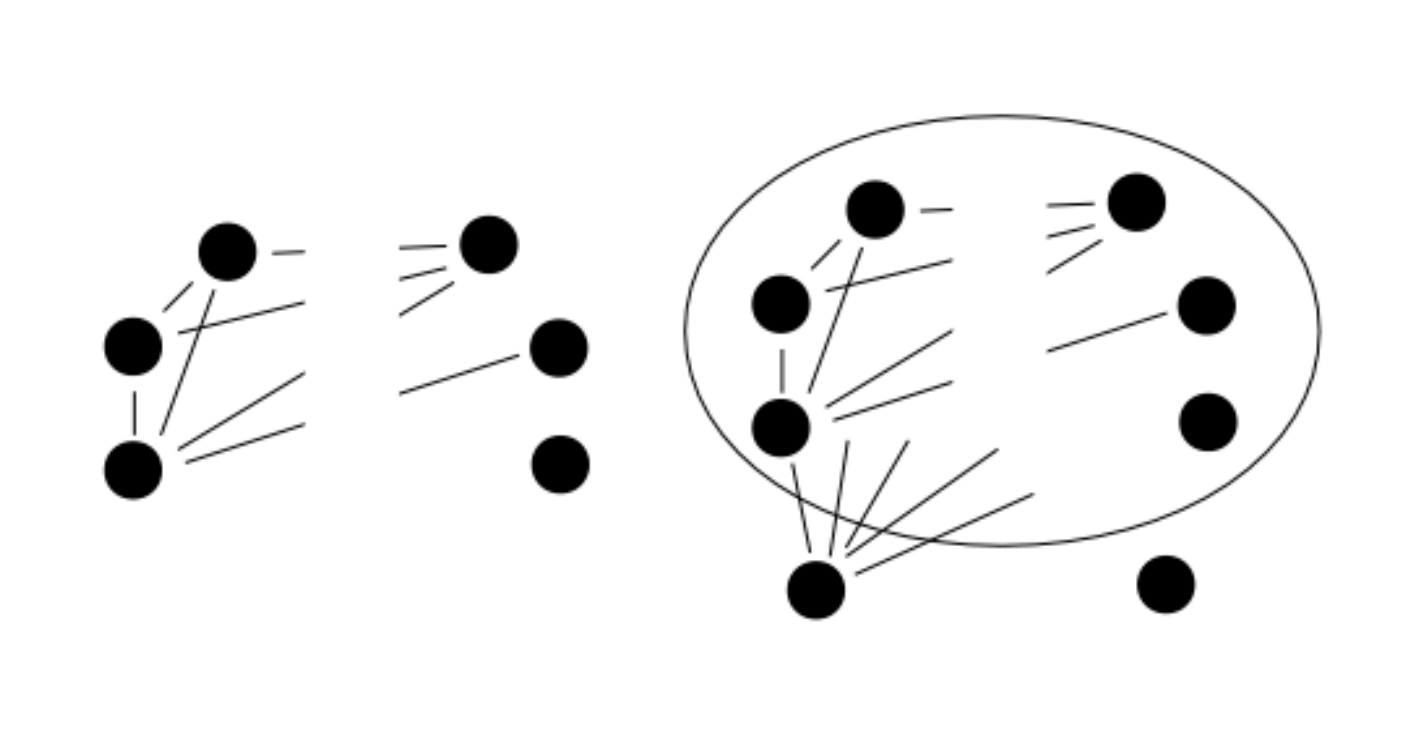}
\caption{\label{FigRec} Left: Sketch of the structure of the matrix $\hat J$ for structure of recursive models for $\ell=1$. Red indicates negative $J_{i,j}$. Right: sketch of the recursion that relates a model with $n$ spins to a model with $n+2$ spins.}
\end{figure}

These matrices can be obtained through a recursive procedure (see Fig. \ref{FigRec} right): we start from $n=2$ with $J_{1,2}=+1$. At each iteration, from a system of $n$ spins, we add two spins, $s_0$ and $s_{n+1}$.
The spin $s_0$ has $J_{0,j}=-1$ with all other spins $j\le n$. The spin $s_{n+1}$ has $J_{i,n+1}=+1$ with all spins, including $s_0$. This generates a matrix with $n+2$ spins with the desired structure\footnote{
For each state at energy $E$ and magnetisation $M=\sum_{i=1}^n s_i$, and each choice of $s_0, s_{n+1}$, we obtain a state of the system with $n+2$ spins with energy
\begin{equation}
\label{eqrec1}
E'=E-J\left[s_0s_{n+1}-s_0 M+s_{n+1}M)\right]
\end{equation}
and magnetisation $M'=M+s_0+s_{n+1}$. 
Thus the energy degeneracy of a system of $n$ spins can be computed as $W_n(E)=\sum_{M}w_{n}(E,M)$,
where $w_n$ satisfies the recursive equation
\begin{equation}
\label{ }
w_{n+2}(E',M')= \sum_{s_0,s_{n+1}=\pm 1} \sum_{E,M} 
w_{n}(E,M)\delta_{E'(E)}\delta_{M',M+s_0+s_{n+1}}.
\end{equation}
Here the shorthand $\delta_{E'(E)}$ reduces the sum to only those terms where Eq. (\ref{eqrec1}) is satisfied.}.

It is possible to generalise the recursion relation to obtain matrices with different values of $\ell$. For example, the case $\ell=n/2$ can be obtained by separating the set of spins into two equal parts, $\mathcal{I}_{\le}=\{i\le n/2\}$ and $\mathcal{I}_{>}=\{i> n/2\}$. To each of the two sets we add two spins at each iteration, thereby obtaining a system with $n+4$ spins. Of the two spins added to $\mathcal{I}_{\le}$, one has $J_{0,j}=-1$ with all spins $j\in\mathcal{I}_{\le}$ and $J_{0,j}=+1$ for all $j\in\mathcal{I}_{>}$. The other three spins have ferromagnetic interactions with all other spins, as well as among themselves. It is easy to see by induction, that the ground state of this model is the ferromagnetic one $s_i=s_j$ for all $i,j$. This holds true for all $\ell\le n/2$.

Numerical iteration of the recursion relations allowed us to compute the degeneracy $W_n(E)$ and the curves $h_E(h_s)$ for different values of $n$ and $\ell$. 

For a given value of $\ell$, there are $\frac{n!}{(n-\ell)!}$ nested models depending on how the set of $n-\ell$ fully ferromagnetic spins are chosen and on the ranking of the $\ell$ remaining spins in the nested hierarchy. This set of models forms a single connected component under a dynamics where a negative link $J_{i,j}=-1$ is ``rewired'' to a different node $k$, i.e. $(J_{i,j},J_{k,j})=(-1,+1)\mapsto (+1,-1)$. In order to show this, let $\partial_i$ be the set of spins with $J_{i,j}=-1$. If we sort spins in such a way that $\delta_1=(1,\ldots,\ell),~\delta_2=(1,\ldots,\ell-1),\ldots,\delta_k=(1,\ldots,\ell-k+1),\ldots $, then the rewiring of the link $(k,\ell-k+1)$ to $(k+1,\ell-k+1)$ interchanges the order of spins $k$ and $k+1$ in the hierarchy. The rewiring of the link $(1,\ell)$ to $(1,k)$ with $k>\ell$ changes the membership of spins in the network connected by $J_{i,j}=-1$. Repeated application of these two moves in the appropriate order makes it possible to reach any model in the set of nested models, from any other model.

\section{RBMs and the reduced MNIST dataset}
\label{AppData}

We focus on the architecture of Restricted Boltzmann Machines (RBM). These have $n_v$ visible  units $\bx=(x_1,\ldots, x_{n_v})$, and $n$ hidden binary units $\bs=(s_1,\ldots,s_n)$ with $s_i=0,1$. The joint probability distribution is 
\begin{equation}
\label{ }
p(\bx,\bs)=\frac{1}{Z}e^{\sum_i b_i s_i+\sum_j c_j x_j+\sum_{i,j}s_i w_{i,j} x_j}
\end{equation}
where $Z=Z(\bb,\bc,\hat w)$ is the partition function. The parameters $\bb,\bc$ and $\hat w$ are learned from the data, in such a way as to maximise the log-likelihood $p(\bx)$. The energy of hidden state $\bs$ is defined as
\begin{equation}
\label{EsRBM}
E_{\bs}=-\log p(\bs)=\log Z-\sum_i b_i s_i-\sum_j \log\left[1+e^{c_j+\sum_{i}s_i w_{i,j}}\right].
\end{equation}
Since $E_{\bs}$ can take any real value, we measure the relevance with respect to a precision $\Delta$ of the energy spectrum, as in Eq.~(\ref{HEDelta}). 

In what follows, we study a reduced MNIST dataset~\cite{MNIST}, where we first coarse grain the original $28\times 28$ pixels data in cells of $2\times 2$ pixels, that are transformed into binary values by applying a threshold\footnote{If the sum of the grey levels of the four pixels exceeds 400, the coarse grained pixel is assigned the value one, otherwise it is zero.}. Finally we discard the first and the last rows and column of cells, thereby leaving us with a dataset of $N=60000$ images of $12\times 12$ binary pixels, corresponding to $n_v=144$. The reason for working with such a reduced dataset is that it allows us to compute the exact distribution of energies, for  values of $n$ in a  relevant range. Indeed, the evaluation of the energy requires the calculation of the partition function $Z$, which can be computed only for small systems.  Sect. \ref{exactRBM} presents results for RBMs with a number of hidden units such that the exact calculation of $p_\Delta(E)$ on all  $2^n$ of states is within reach of our computational resources. 
For larger systems we sample $N\ll 2^n$ states $\bs$ from $p(\bs)$ with MCMC method and estimate $H_\Delta(E)$ from the sample (see Sect.~\ref{sampleRBM}). 


In order to test whether learning machines converge to models of maximal relevance during training, we focus on the architecture of Restricted Boltzmann Machines (RBM). These have $n_v$ visible  units $\bx=(x_1,\ldots, x_{n_v})$, and $n$ hidden binary units $\bs=(s_1,\ldots,s_n)$ with $s_i=0,1$. The joint probability distribution is 
\begin{equation}
\label{ }
p(\bx,\bs)=\frac{1}{Z}e^{\sum_i b_i s_i+\sum_j c_j x_j+\sum_{i,j}s_i w_{i,j} x_j}
\end{equation}
where $Z$ is the partition function, which depends on the parameters $\bb,\bc$ and $\hat w$ that are learned from the data, in such a way as to maximise the log-likelihood $p(\bx)$. The energy of hidden state $\bs$ is defined as
\begin{equation}
\label{EsRBM}
E_{\bs}=-\log p(\bs)=\log Z-\sum_i b_i s_i-\sum_j \log\left[1+e^{c_j+\sum_{i}s_i w_{i,j}}\right].
\end{equation}
Since $E_{\bs}$ can take any real value, we measure the relevance with respect to a precision $\Delta$ of the energy spectrum, as in Eq.~(\ref{HEDelta}). 

\bibliographystyle{ieeetr}
\bibliography{biblio.bib}

\end{document}